\documentclass{article}

\usepackage[english]{babel}

\usepackage[letterpaper,top=2cm,bottom=2cm,left=3cm,right=3cm,marginparwidth=1.75cm]{geometry}

\usepackage{amsmath}
\usepackage{graphicx}
\usepackage[colorlinks=true, allcolors=blue]{hyperref}

\title{Space-Time Encoded Modulation for High-Fidelity Diffuse Optical Imaging}
\author{Ben Wiesel, Shlomi Arnon}

\begin{document}
\maketitle

\begin{abstract}
Diffuse optical imaging (DOI) offers valuable insights into scattering mediums, but the quest for high-resolution imaging often requires dense sampling strategies, leading to higher imaging errors and lengthy acquisition times. This work introduces Space-Time Encoded Modulation (STEM), a novel light modulation scheme enabling low-noise, high-resolution imaging with single-pixel detectors. In STEM, a laser illuminates the sample, and the transmitted light is detected using a single pixel detector. The detected image is partitioned into a two-dimensional array of sub-images, each encoded with a unique quasi-orthogonal code. These coded sub-images represent light transmission at specific locations along the sample boundary. A single-pixel detector then measures their combined transmission. By virtue of their quasi-orthogonality, the relative strength of each sub-image can be measured, enabling image formation. In this paper, we present a comprehensive mathematical description and experimental validation of the STEM method. Compared to traditional raster scanning, STEM significantly enhances imaging quality, reducing imaging errors by up to 60\% and yielding a 3.5-fold increase in reconstruction contrast.
\end{abstract}

\section{Introduction}

Looking through and inside turbid medium using visible, or near-infrared (NIR) light holds great significance across a wide range of scientific disciplines [1]. As light passes a scattering medium it undergoes multiple scattering events. Driven by its inherent stochastic nature, photons diverge from their original trajectories and follow a complicated random path, making image retrieval challenging [1]. One of the most promising techniques developed in this context is diffuse optical imaging (DOI) [1]–[3]. In DOI, an array of light sources illuminates the tissue, and the scattered light is measured with an array of detectors. Then, the reconstruction is done using model-based algorithms, where one attempts to “invert” a propagation model to image the object of interest [1]. DOI allows to image centimeters into objects with mm scale resolution and has been deployed in various potential applications such as breast [4] and brain imaging [5]. Despite DOI's considerable achievements, its practical implementation encounters substantial hurdles. The fundamental reconstruction challenge it faces is inherently ill-posed and nonlinear, resulting in most algorithms producing low-resolution images riddled with noise and artifacts.
	Enhancing resolution and minimizing artifacts has therefore been the focus of prior investigations, highlighting the significance of increasing the amount of data extracted from the sample. This objective can be accomplished through the incorporation of either spectral [6], [7] or temporal [3], [8] data, and by implementing dense scanning strategies [6], [9][10], [11]. However, these approaches often require advanced single-pixel detectors and meticulous raster scanning, leading to extremely long acquisition and computational times, hindering their practical implementation in clinical settings. While methods that enable to acquire spatial information without compromising detection time are available, they typically incur significant cost and are limited in their achievable resolution [12], [13].  
Single-pixel imaging emerges as a powerful solution, offering high-resolution images with high dynamic range at a fraction of the cost of traditional methods [14], [15]. Additionally, recent advancements in ultra-fast spatial light modulation [16], [17] open doors for developing novel imaging methods and stands as a pivotal component in our quest to push the boundaries of real-time diffuse imaging.
This work introduces a novel single-pixel imaging approach, Space-Time Encoded Modulation (STEM), a novel light modulation scheme that leverages the capabilities of SLMs to achieve low-noise, high-resolution DOI imaging with single-pixel detectors. Unlike previous methods that utilized spatial modulation of light [14], [18], [19] our .
method introduces temporal modulation for the first time. Which involves different families of modulation patterns to be used and novel image reconstruction algorithms to be developed. STEM draws inspiration from the principles of code division multiple access (CDMA) [20], [21] which enables multiple users to share a common channel by encoding their data with unique signature codes. By carefully designing the codes to have good correlation properties, the inter-user interference can be minimized, and each user’s signal can be accurately read [22], [23]. This same concept can be applied to diffuse imaging, where the primary goal is typically to measure the scattering of light with various detectors distributed throughout the sample. In the transmission configuration illustrated in Figure 1, this process entails illuminating the sample with a wide light source and capturing the transmitted light image using a single-pixel device. This involves several steps: First, the transmitted light is divided into a 2D array of sub-images, where each sub-image is being modulated by a unique signature code using a spatial light modulator (SLM). Subsequently, a single-pixel device records the combined intensity of all sub-images during their modulation period. Finally, to form an image, each sub-image’s allocated code is correlated with the detected signal to measure its relative intensity.
This paper contributions can be summarized as follows: 1) The first application of temporally modulated light patterns with orthogonality in time for single-pixel imaging. 2) Development of a novel single-pixel imaging reconstruction algorithm based on temporal correlations. 3) Formulation of a comprehensive mathematical model enabling optimization of illumination patterns and a priori assessment of the modulation scheme effectiveness. 4) Introduction of a temporal compression technique which allows for reduced acquisition times while preserving most spatial information.

\begin{figure}
\centering
\includegraphics[width=1\linewidth]{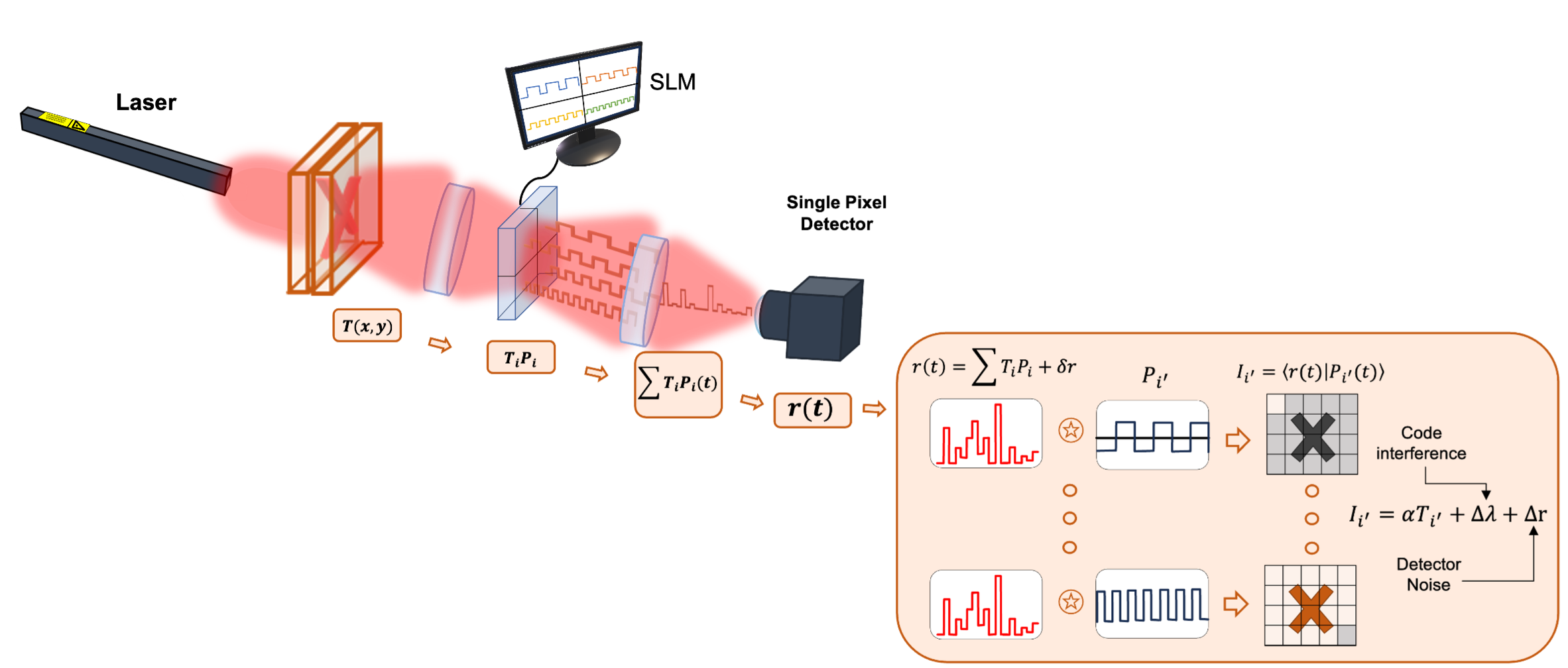}
\caption{\label{fig:Fig1} Transmission configuration STEM: A wide laser source illuminates the sample, and the total transmitted light is measured using a single-pixel device. To form an image, the transmitted light is divided into a 2D array of sub images, each modulated with a unique signature code using an SLM. An image can be formed by correlating each sub image’s assigned code with the detected signal. Two primary noise mechanisms affect image quality: code interference and detector noise.}
\end{figure}

\section{Methods}

\subsection{STEM in a nutshell}

The experimental setup (Figure 1) uses a continuous wave laser to illuminate a sample, to measure the spatially varying transmittance \( T \) via single-pixel detection. Here, \( T \) is a 2D image comprised of \( N \) unknown intensities \( T_i \) at pixel \( i \) (\( N = N_x \cdot N_y \)). A lens relays the transmittance image onto the SLM plane, where pixel-dependent binary signature codes modulate the light intensity. Finally, by focusing the light, a detector measures the combined intensity from all coded modulations. The signal at the output of the detector can be written as follows [24]:

\begin{equation}
r(t) = \sum_{i=1}^{N} T_i P_i(t) + \delta r(t)
\end{equation}

Where \( \delta r(t) \) is the detector noise and \( P_i(t) \) is the \( i \)-th pixel assigned signature code out of a code family \( C \). An \( (n, w, \lambda_m, \delta \lambda) \) signature code is a family of \( (0,1) \) sequences of length \( n \), and weight \( w \). The code weight \( w \) is the number of ones in the sequence and remains constant across all codes in \( C \). The code size \( |C| \) is the number of sequences in the family. The codes are projected onto the SLM as binary on/off keys \( A_{i,j} \) in frames of duration \( T_c \), given by [25]:

\begin{equation}
P_i(t) = \sum_{j=0}^{n-1} A_{i,j} P_{T_c}(t - jT_c)
\end{equation}

Where \( P_{T_c} \) is the rectangular pulse of duration \( T_c \), and the total measurement time is denoted as \( T \) (i.e., \( T = n \cdot T_c \)).

The signature codes’ correlation definitions have undergone several changes compared to the traditional definitions used in multiple access applications [20], [21]. First, the constraint on the signature code’s autocorrelation is removed completely since we only deal with synchronous signals. Second, a new parameter is defined: the mean pairwise cross correlation across all code pairs, \( \lambda_m \). Mathematically, it can be written as:

\begin{equation}
\lambda_m = \frac{1}{|C|^2 - |C|} \sum_{i=1}^{|C|} \sum_{\substack{i' \neq i}} \sum_{j=0}^{n-1} A_{i,j} A_{i',j}
\end{equation}

Where \( A_{i,j}, A_{i',j} \) are the \( j \)-th entries from two different codes in \( C \), and the coefficient \( \frac{1}{|C|^2 - |C|} \) is the number of non-identical code pairs. Unlike the traditional bound on the maximal pairwise cross correlation [20], [21], we define here a new bound \( \delta \lambda \), which is the maximal cross-correlation shift from its mean \( \lambda_m \) across all code pairs. Using (3), this bound can be written as:

\begin{align}
\sum_{j=0}^{n-1} A_{i,j} A_{i',j} &= \lambda_m + \delta \lambda_{i,i'} \quad ; \quad \forall i \neq i' \\
|\delta \lambda_{i,i'}| &\leq \delta \lambda \quad ; \quad \forall i \neq i'
\end{align}

Where \( \delta \lambda_{i,i'} \) is the specific correlation value shift from \( \lambda_m \), which can take any positive or negative value. Using Eq. (2) and the code weight and cross-correlation definition from (4), the correlation of two signature sequences integrated over the total measurement time \( T \) is given by:

\begin{equation}
\langle P_i \mid P_{i'} \rangle = T_c \cdot
\begin{cases}
w & \text{if } i = i' \\
\lambda_m + \delta \lambda_{i,i'} & \text{if } i \neq i'
\end{cases}
\end{equation}

To attain the image and specifically to reconstruct a given pixel intensity value, its corresponding code is correlated with the detected signal. For pixel \( i' \), we correlate the received signal \( r(t) \) with \( P_{i'}(t) \). Using (1), we attain:

\begin{equation}
I_{i'} = \langle r \mid P_{i'} \rangle = \sum_{i=1}^{N} T_i \langle P_i \mid P_{i'} \rangle + \langle \delta r \mid P_{i'} \rangle
\end{equation}

By defining \( M \) as the total transmittance projected on the SLM (i.e., \( M = \sum_i T_i \)) and further simplifying this relation, we identify several key observations (the full mathematical derivation can be found in Supplementary Material, Section 2.2). First, the total signal is composed of four terms: the true signal, a bias term, and two noise terms.

\begin{equation}
I_{i'} = T_c (w - \lambda_m) T_{i'} + M T_c \lambda_m + \Delta \lambda + \Delta r
\end{equation}

The bias term does not affect image quality since it can always be removed by measuring \( M \) experimentally or simply by rescaling the attained image. 

The first noise term (interference noise \( \Delta \lambda \)) is related to the cross-correlation shift \( \delta \lambda \) defined above. Its associated noise variance is given by:

\begin{equation}
\sigma_\lambda^2 = \text{Var}[\Delta \lambda] = \left( T_c \delta \lambda (M - T_{i'}) \right)^2
\end{equation}

It is important to emphasize that the non-zero cross-correlation (\( \lambda_m \neq 0 \)) is not the cause of this interference noise. The actual mechanism behind it is the non-uniformity in the pairwise correlation (\( \delta \lambda \neq 0 \)).

The second noise term (measurement noise \( \Delta r \)) arises from the detector noise, primarily thermal noise, shot noise, and dark current noise. Following previous works assumptions [26]–[28], the system noise is modeled as additive white Gaussian noise (AWGN) with zero mean and variance \( \sigma^2 \), and is given by:

\begin{equation}
\sigma_r^2 = \text{Var}[\Delta r] = \sigma^2 T_c w
\end{equation}

To understand the individual effect of each noise source and their implication for designing optimal codes, we distinguish between two cases. The first case is when the interference noise is much larger than the measurement noise ($\sigma_\lambda^2 \gg \sigma_r^2$). This can happen for code families with relatively high correlation shifts. For this case, we attain the following SNR:

\begin{equation}
\text{SNR}_{\Delta \lambda} \approx \frac{T_c (w - \lambda_m) T_{i'}}{T_c \delta \lambda (M - T_{i'})} \propto \frac{w - \lambda_m}{\delta \lambda}
\end{equation}

We aim to identify SNR dependencies that are not image specific. Hence, we will disregard disregard the \( T_{i'} / (M - T_{i'}) \) ratio. The relation offers valuable insights for designing and choosing optimal code families. We observe a linear relationship of the SNR with the disparity between \( w \) and \( \lambda_m \), and an inverse relationship with \( \delta \lambda \). Therefore, codes that maintain uniform cross-correlation (\( \delta \lambda = 0 \)) are resilient to interference noise. The second case is when the measurement noise is much larger than the interference noise  ($\sigma_r^2 \gg \sigma_\lambda^2$), which can happen for highly noisy channels. For this case we attain the following SNR:

\begin{equation}
\text{SNR}_{\delta r} = \frac{T_c (w - \lambda_m) T_{i'}}{\sigma \sqrt{T_c w}} = \frac{\sqrt{T_c} (w - \lambda_m) T_{i'}}{\sigma \sqrt{w}}
\end{equation}

Once again, this quantity exhibits proportionality to the difference between \( w \) and \( \lambda_m \). This reaffirms that, despite the earlier observation, having a low \( \lambda_m \) is also necessary to enhance robustness against measurement noise. Moreover, we uncover two other intuitive relationships. First, the SNR displays linear proportionality to the light intensity \( T_{i'} \), aligning with expectations from traditional imaging. Second, the SNR is linearly related to the square root of the frame time \( T_c \), mirroring the square root improvement seen in conventional imaging with increasing integration time. Lastly, we observe an inverse relationship between the SNR and the square root of the weight. This underscores the significance of using code families with a maximized \( \frac{w - \lambda_m}{\sqrt{w}} \) ratio for optimal performance.

\subsection{Signature code construction}

This article explores three distinct illumination methodologies for STEM. Each methodology exhibits unique characteristics conducive to optimal performance under varying conditions. We investigate how these attributes manifest in real-world scenarios, analyzing their strengths and limitations. The specific properties and construction rules of each methodology are outlined below, followed by a comparative analysis of their respective attributes in Table 1.

\textbf{Raster Encoding (RE)}: The first approach, Raster Encoding (also known as raster scanning) [29], relies on a simple but effective illumination strategy. In each frame, only one pixel on the SLM plane is activated ("ON"), effectively simulating sequential scanning. This translates to employing signature codes with mostly zeros ("OFF") and a single "ON" element positioned at the index corresponding to the code's location within its family. RE scanning therefore has a code weight $w=1$, mean correlation $\lambda_m=0$, correlation maximal shift $\delta\lambda=0$ and can accommodate any arbitrary number of pixels.

\textbf{Random Optical Encoding (ROE)}: ROEs [21] represent a family of binary codes consisting of randomly positioned zeros and ones (0,1). The code weights were kept constant across sequences. This was done using several steps. First, all $N$ signature codes were initialized to the zero vector of length $n$. Second, $N$ unique random permutations of the integers $1$ to $n$ were chosen. Finally, the first $w$’th entries of each permutation were chosen and their value was replaced from $0$ to $1$. Concluding in a total of $N$ distinct random code with $w=n/2$. The average cross correlation for this case can therefore be estimated to be $\lambda_m=n/4$. However, because the placement of ones (1s) in ROE is entirely random, the correlation properties will vary between code pairs and a non-zero correlation shift will be apparent ($\delta\lambda>0$).

\textbf{Hadamard Encoding (HE)}: A Hadamard code [21] of length $n$ is generated using rows from an $n \times n$ orthogonal Hadamard matrix $H_M$ with binary values $(-1,1)$ where $n=2^M, M\in{N}$. However, since light intensity is a non-negative phenomenon, only unipolar codes $(0,1)$ can be used. Therefore, a modified unipolar Hadamard matrix was used, which can be attained by replacing the $-1$s with $0$s. For example, the unipolar Hadamard code for $n=4$ is given by:

\[
H_2 =
\begin{bmatrix}
1 & 1 & 1 & 1 \\
1 & 0 & 1 & 0 \\
1 & 1 & 0 & 0 \\
0 & 0 & 0 & 1
\end{bmatrix}
\tag{13}
\]

HE has a code weight $w=n/2$, mean correlation $\lambda_m=n/4$, correlation maximal shift $\delta\lambda=0$ and can accommodate only $n=2^M$ users.

\begin{table}[h]
\centering
\caption{Illumination Codes Comparison}
\vspace{0.5em} 
\begin{tabular}{c|c|c|c|c}
\textbf{} & {Code Size ($|\mathcal{C}|$)} & {Code Weight ($w$)} & {Mean Correlation ($\lambda_m$)} & {Correlation Shift ($\delta\lambda$)} \\
\hline
RE  & $|\mathcal{C}| \in {N}$           & 1         & 0             & 0 \\
ROE & $|\mathcal{C}| \in {N}$           & $n/2$     & $n/4$         & $>0$ \\
HE  & $|\mathcal{C}| = 2^M,\ M \in {N}$ & $n/2$     & $n/4$         & 0 \\
\end{tabular}
\end{table}

\subsection{Experimental setup}

Our objective is to establish a setup capable of effectively employing the STEM technique. This configuration must possess the ability to generate spatiotemporal light masks with high precision, consistency, and with detectable modulations. The experimental arrangement for the transmission configuration is depicted in Figure 2. Here, we utilize an expanded laser beam at 640 nm, emitting 50 mW of power (Becker \& Hickl, BDS-SM-640), to illuminate an object sandwiched between two scattering slabs. Subsequently, the scattered light, upon passing through the medium, is directed onto a transmissive spatial light modulator (Holoeye, LC 2012) equipped with pixel-based temporal modulations. This spatial light modulator (SLM) serves the purpose of partitioning the transmission image into an array of sub-images, where each is modulated with a distinct signature code. Finally, the total modulated light is focused onto a wide bandwidth optical receiver (Thorlabs, PDA36A), capable of detecting light signals across a wavelength range of 350 to 1,100 nm. Following the detection phase, image formation occurs during postprocessing, involving the correlation of each pixel's assigned code with the detected signal to determine its relative power.

\begin{figure}
\centering
\includegraphics[width=1\linewidth]{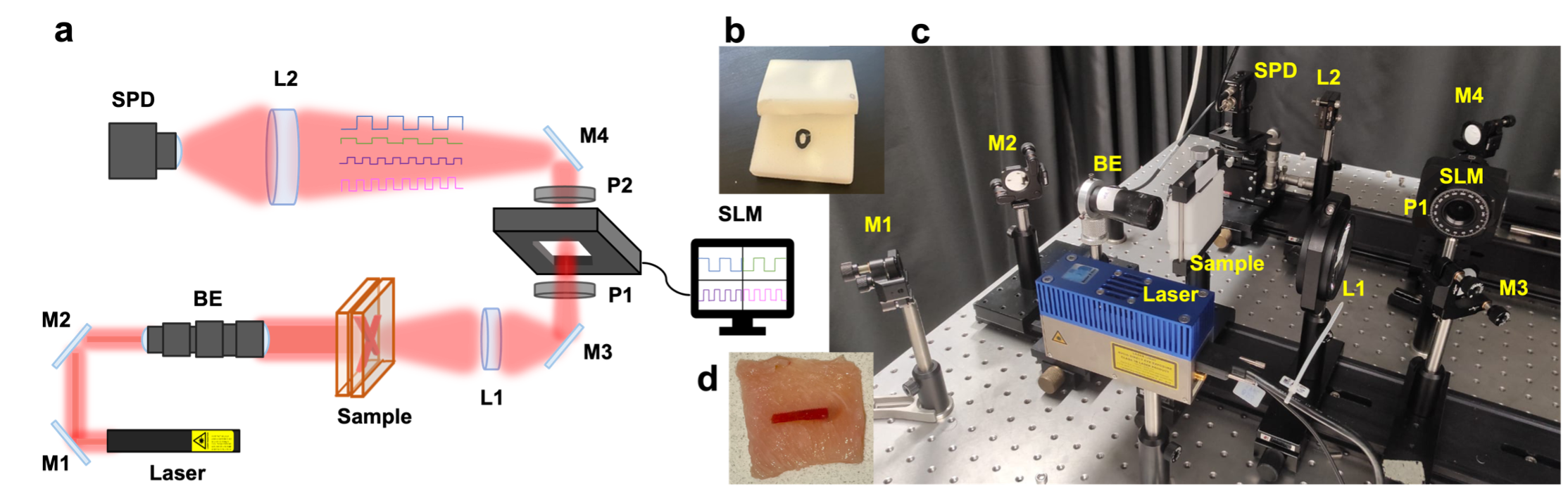}
\caption{\label{fig:Fig2} STEM schematic setup: a, a CW laser is directed (M1+M2) towards a beam-expander (BE) and projected onto the sample. After passing the sample, the total transmitted light is imaged (L1) and directed (M3) onto the SLM, which is sandwiched between two linear polarizers (P1+P2) to generate the spatiotemporal modulations. Finally, the modulated light is directed (M4) and imaged (L2) onto the optical receiver. The total transmitted light is composed of a summation of different modulation components, which can be utilized to form an image using correlation analysis.  b, the absorber is an alphanumerical black stamp. c, experimental setup. d, the ex-vivo phantom with embedded anomaly.}
\end{figure}

\subsection{Phantom preparation}

Two distinct scattering phantoms were employed to evaluate the system's performance under varying conditions. The first phantom utilized melamine foam (Figure~2b), a well-established scattering medium [8], [30]. This phantom consisted of two 1~cm melamine slabs with embedded black plastic letter stamps to introduce high absorption contrast. The absorption and reduced scattering coefficients of the melamine foam were determined using single-point, time-resolved measurements to be $\mu_a = 0.02~\mathrm{cm}^{-1}$ and $\mu_s = 8.28~\mathrm{cm}^{-1}$ (see Supplementary Material, Section 2.4).

A second, biological phantom was prepared using fresh chicken breast (Figure~2d). The chicken breast was carefully sliced to a uniform thickness of approximately 7~mm, with the skin and fat removed, using a professional food slicer. The slices were then warmed to room temperature to simulate physiological conditions and ensure consistency in optical properties during the experiments. To simulate anomalies, beef semimembranosus muscle sections were prepared and aged for 17 days at $3^\circ$C. After aging, a small section of approximately $4 \times 20 \times 2~\mathrm{mm}^3$ was made using a precision knife and inserted between two chicken breast slabs at room temperature. This “sandwiched” configuration, which ensured a consistent inclusion depth, is illustrated in Figure~4C and D. The preparation followed the protocol from [31] to ensure consistency and reproducibility. Semimembranosus was selected due to its known optical properties and higher absorption properties compared with the chicken breast. The optical properties at 640~nm for the chicken breast medium are taken from literature [32] to be $\mu_a = 0.17~\mathrm{cm}^{-1}$ and $\mu_s = 2.7~\mathrm{cm}^{-1}$, and $\mu_a = 0.5~\mathrm{cm}^{-1}$ and $\mu_s = 6.2~\mathrm{cm}^{-1}$ for the semimembranosus muscle [31].

\section{EXPERIMENTS}

\subsection{Low-light image acquisition using STEM }

Our objective in this section is to assess the imaging capabilities of STEM in real-world scenarios, particularly under low-light conditions. For each captured image, we also obtained a reference image using a CMOS camera beam profiler (Thorlabs, BC207VIS(/M)). This reference image serves as our gold standard for subsequent comparisons. Throughout the paper we refer to the grid size as $N$, which is the number of pixels in the image. In most instances, the code length $n$, signifying the number of frames in the sequence, matches the grid size (i.e., $n = N$). However, in some cases the code length can be larger ($n > N$), which will increase the noise robustness, or can be even lower ($n < N$), as a form of compressive sampling at a cost of reducing the image quality. The methods are compared in terms of pixel-wise root-mean-square-error (RMSE) or the peak-signal-to-noise-ratio (PSNR) of the reconstructed image. Figure~3 presents the comparison in terms of RMSE as a function of the pixel number (Figure~3a), and the PSNR as a function of the integration time (Figure~3b). 

To further validate the effectiveness of STEM, we applied several denoising methods to the RE measurements. This analysis enabled a comparative evaluation of STEM against other approaches for enhancing imaging SNR. The two most effective methods are illustrated in Figure~3. The first method, {RE-Gaussian}, applies a two-dimensional Gaussian filter to the detected image from the standard RE pipeline. The second method, {RE-Tik 2nd}, employs Tikhonov regularized inversion, which addresses ill-posed problems by introducing a penalty term $\lambda \|Lx\|_2^2$ to the least-squares objective, where $L$ is the second-order difference operator that imposes smoothness on $x$. A detailed Analysis of all denoising methods and their comparison can be found in the Supplementary Material, Section 2.7. Increasing the integration time involved using longer sequences for the same number of pixels ($n > N$), this way each sequence weight is increased which as explained improves the reconstruction SNR (see Supplementary Material, Section 2.5). Several observations should be noted. First, the RMSE increases with increasing grid size for both the RE and HE methods, which aligns perfectly with our mathematical model. As explained above, both RE and HE exhibit zero correlation shift $\delta\lambda$, making them immune to interference noise. Nevertheless, they are still subject to the effects of measurement noise $\Delta r$. From Eq. 12, we find that $\text{SNR}_{\delta r}$ depends on the optical power at each pixel on the SLM plane ($T_i$). Consequently, when the grid size increases, the average optical power on each pixel must decrease leading to a lower $\text{SNR}_{\delta r}$ ratio and a higher RMSE. Furthermore, it becomes apparent that the RMSE for the RE method increases at a higher rate than that of the HE approach. This is explained by the opposed linear dependence of the $\text{SNR}_{\delta r}$ on $w - \lambda_m$ which remains constant for the RE method but increases with the grid size ($N$) for the HE approach. 

\begin{figure}
\centering
\includegraphics[width=1\linewidth]{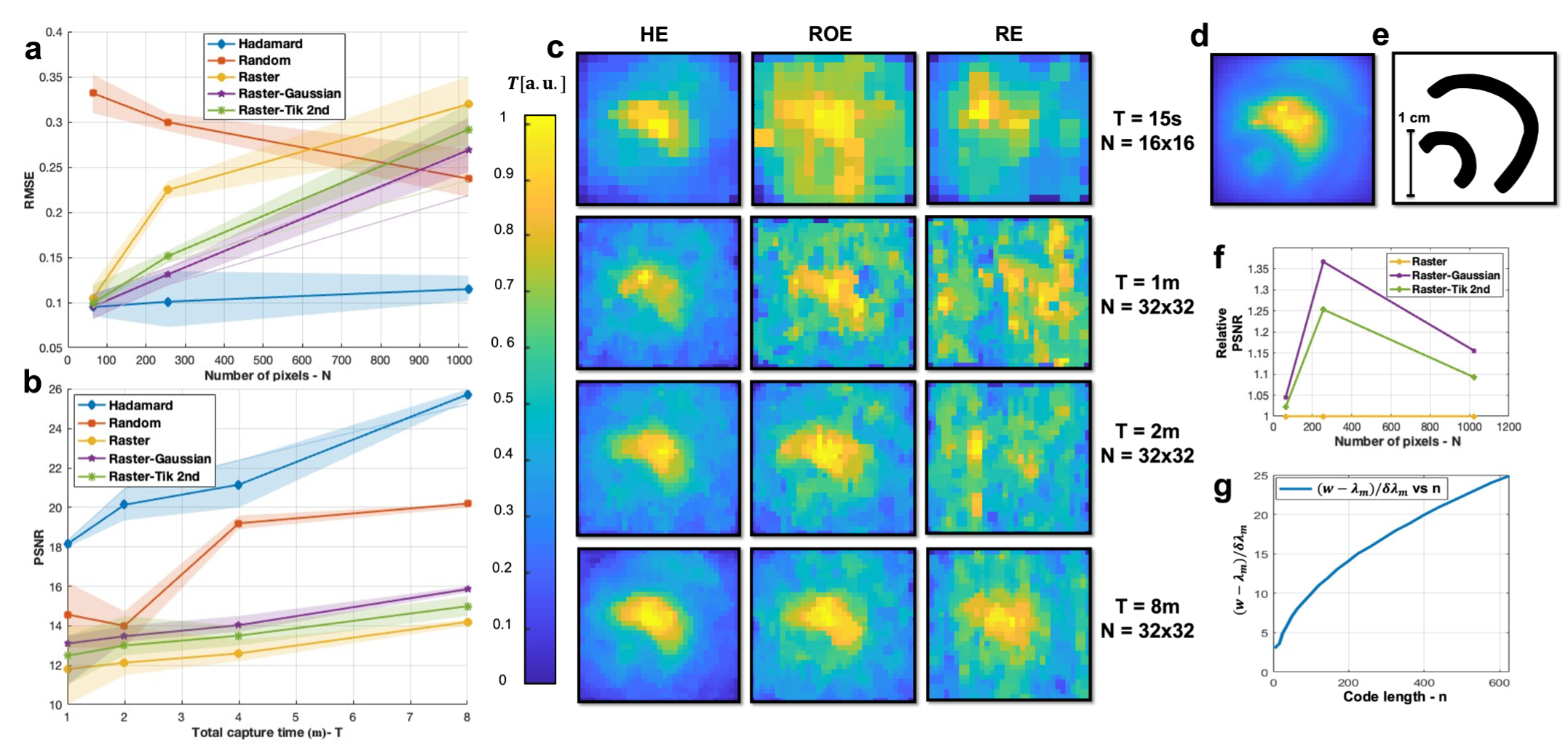}
\caption{\label{fig:Fig3} STEM experimental results for low light image capture. a, pixel wised RMSE vs the pixel grid size for the different illumination methodologies. HE provides a better RMSE across all configurations. b, pixel-wise PSNR vs. total integration time (T) for a fixed grid size of 32×32 (N=1024) using the different methodologies. Increasing T leads to better SNR ratios and increases the noise robustness. HE allows for higher PSNR and better enhancement rate than competing methods. c, measured light intensity image samples reconstructed using STEM for different illumination methods for varying integration times and grid sizes. d, the ground truth low-light image captured with a CMOS beam profiler, serves as our gold standard for comparisons. e, the ground truth anomaly embedded inside the media. f, PSNR of each denoising method, normalized by (relative to) the regular raster results. g, graph of (w-$\lambda_m$)/$\delta\lambda$ vs code length. This observation implies that interference robustness increases with the grid size for ROE.}
\end{figure}

The second observation pertains to the ROE case and reveals two notable distinctions. Firstly, there is a notably high RMSE even at smaller grid sizes. Secondly, there is a noticeable decreasing trend in RMSE as the grid size increases, which diverges from our earlier findings. Our explanation for these effects is that the significant noise term for ROE is the interference noise ($\sigma_{\delta\lambda}^2 \gg \sigma_r^2$). Through simulations we were able to prove that for ROE, the interference noise SNR ($\text{SNR}_{\delta\lambda}$) increases for larger code lengths $n$ (Figure 3f, for further details see Supplementary Material, Section 2.6). Hence, since $\Delta\lambda$ is the dominant noise term, we experience a large error for small grid size (small $n$) which decreases as the grid size expands.

Another observation is the higher improvement rate of the image PSNR of the HE approach for increasing capture times $T$ (Figure 3b). By observing (23) it can be noted that when $T$ is increased, the signature code weights for HE increase faster than the RE approach, therefore, $\text{SNR}_{\delta r}$ increases faster as well.

Finally, comparisons of HE and ROE with denoised RE approaches reveal several trends. As depicted in Figure 3, denoising improves the imaging SNR across all configurations. However, HE and ROE consistently outperform the denoised RE, particularly for larger grid sizes. This result arises from the inherently higher theoretical SNR associated with HE and ROE. It is also worth noting that the SNR gain from denoising exhibits a non-monotonic trend: it initially increases and then diminishes as the grid size grows (Figure 3f). For small grids, the baseline SNR for RE is already high, limiting the denoising gain. Conversely, for large grids, the baseline SNR becomes too low for denoising to recover the lost information effectively. These findings collectively underscore the effectiveness of the HE method, which exhibit both low correlation shift $\delta\lambda$ and high w to $\lambda_m$ difference leading to an improved performance and to faster error reduction rates.

\begin{figure}
\centering
\includegraphics[width=1\linewidth]{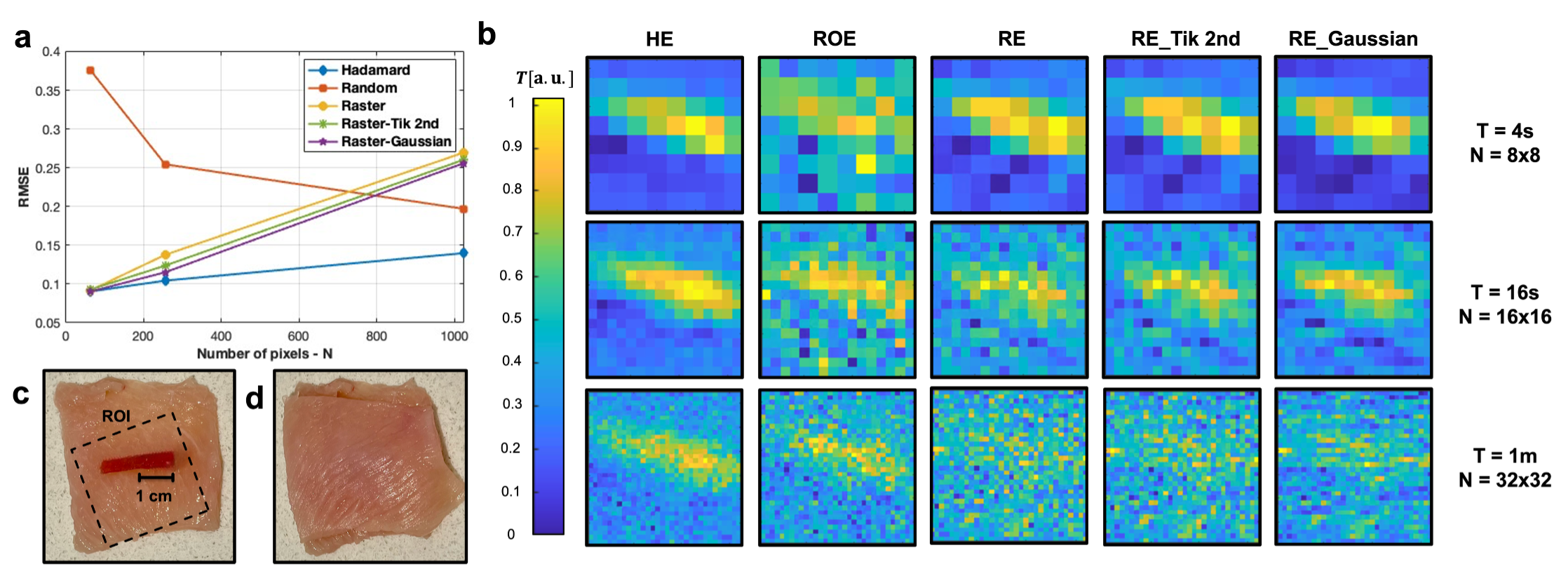}
\caption{\label{fig:Fig4} STEM experimental results for ex-vivo low light imaging. a, pixel wised RMSE vs the pixel grid size for the different illumination methodologies. b, measured light intensity image samples reconstructed using STEM for the different methods for varying grid sizes.  c, the embedded object, with the imaging ROI marked in dotted line. d, the full ex-vivo phantom with the embedded object sandwiched inside.}
\end{figure}

\subsection{Ex-vivo low-light experiments}

To further validate the effectiveness of STEM, we extended our experiments to an ex-vivo setup. The scattering medium consisted of chicken breast tissue embedded with small sections of beef Semimembranosus. This configuration introduced realistic challenges, such as heterogeneous optical properties, while providing a realistic absorption contrast between the embedded targets and the surrounding medium. Figure 4 illustrates the imaging RMSE as a function of grid size N and shows similar trends to the ones observed in previous section. First, again, HE consistently achieved the lowest RMSE across all grid sizes, reaffirming its robustness to noise even under complex scattering conditions. Second, while RE performs comparable to HE at lower grid sizes, its lower theoretical SNR extends for the ex-vivo measurements as well and a sharp increase in imaging error is observed as the grid size grew. Denoising methods moderately reduced the RMSE for RE, particularly at smaller grid sizes. However, even with denoising, RE's performance remained significantly inferior to HE as grid size increased. Finally, ROE followed the trend observed in the previous experiment, with RMSE decreasing as the grid size increased due to reduced interference noise with longer code lengths. However, even at higher grid sizes, ROE was outperformed by HE. These results highlight STEM’s potential to address challenges in optical imaging with high fidelity and efficiency even under real-world conditions.

\begin{figure}
\centering
\includegraphics[width=1\linewidth]{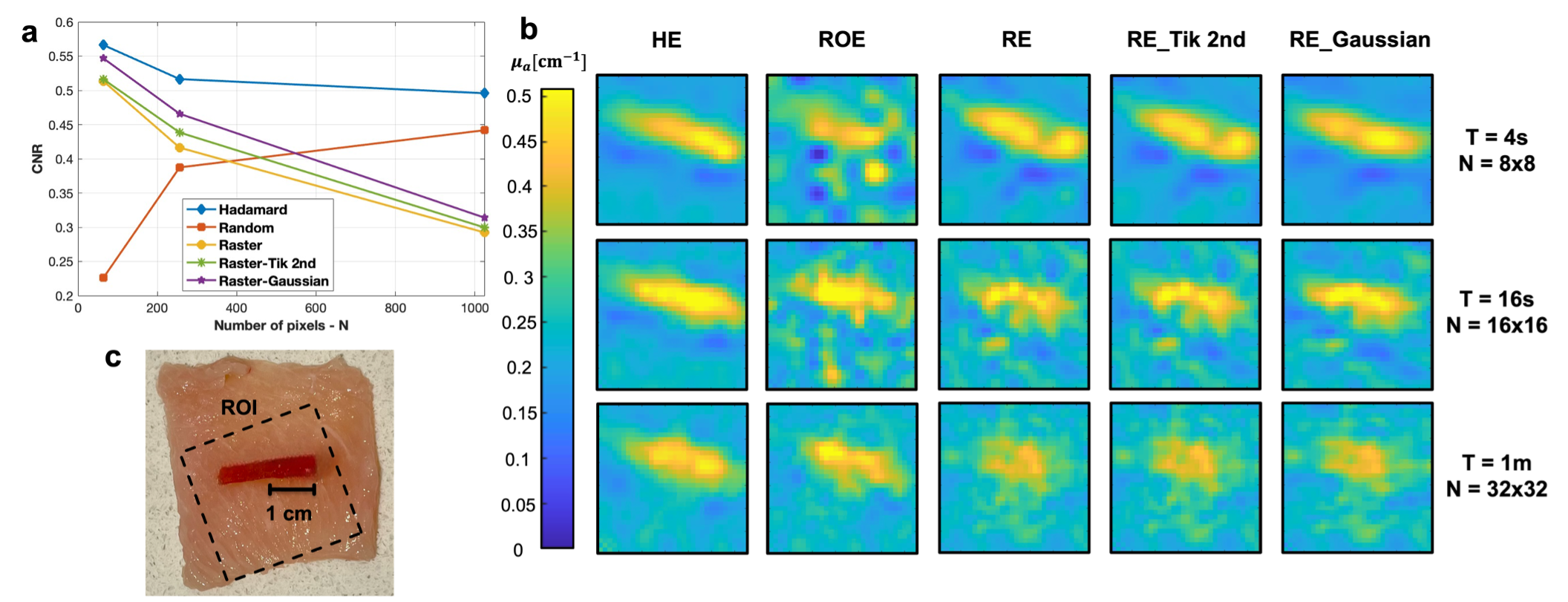}
\caption{\label{fig:Fig5} STEM experimental results for diffuse imaging. a, contrast-noise-ratio (CNR) vs the pixel grid size for the different illumination methodologies. HE consistently outperforms its counterparts in terms of anomaly contrast for all grid sizes. b, anomaly image samples reconstructed using DOI for the different methods for all grid sizes. c, the embedded object, with the imaging ROI marked in dotted line.}
\end{figure}

\subsection{STEM based diffuse optical imaging }

Our ultimate objective is using STEM for DOI, which involves using the acquired images to reconstruct an embedded anomaly. Image reconstruction is attained by minimizing the following formula:

\[
\min_{\mu} \| m - f(\mu) \| + \Lambda(\mu) \tag{14}
\]

Where, $\mu$ is the optical properties distribution of the whole sample; $m$ is the collected measurements which describe the light intensity incident on the detectors; $f(\cdot)$ is the forward model, which connects between the optical properties and the measured intensity; and $\Lambda(\mu)$ is an appropriately chosen regularization term.

By using a linearized reconstruction, the functional relationship from Eq. 14 is approximated as a matrix multiplication $f(\mu) \approx H\mu$, where $\mu$ now represents the change in the optical properties due to heterogeneities in the media (i.e., the embedded object), and $H$ is the measurement sensitivity matrix [33]. To estimate $H$, Monte-Carlo-based simulations were used, since they offer the most versatility and accuracy for our purposes [33]. To estimate $H$, we treat each pixel on the SLM as an individual single detector, and the sensitivity matrix is calculated with respect to a detector placed at the SLM’s pixel location.

The modified objective function is written as follows:

\[
\min_{\mu} \| m - H\mu \| + \lambda_1 \| \mu \|_2 + \lambda_2 \| \mu \|_{\mathrm{TV}} \tag{15}
\]

Where $\| \mu \|_2$ and $\| \mu \|_{\mathrm{TV}}$ are the L2 norm and TV norm [34], and $\lambda_1, \lambda_2$ are their respective tuning parameters (a detailed comparative study on the individual and combined effects of the regularization terms can be found in the Supplementary Material Section 2.9).

Our work uses recent advancements in auto-differentiation to calculate the gradient of the objective function and to solve it efficiently. PyTorch was used to implement our optimization problem, and ADAM was chosen as the optimizer [35].

Figure 5 presents the comparison of the reconstructed images in terms of contrast-to-noise ratio [36] (CNR) for the different illumination methodologies. CNR is used to define the contrast level of the anomaly compared to its surrounding background where higher CNR corresponds to better anomaly detection. The full details for calculating the CNR can be found in its original article [36]. It's important to emphasize that when we increase the size of the imaging grid, two contrasting effects come into play. On one hand, as demonstrated in the previous section, enlarging the grid size reduces the imaging SNR, consequently leading to increased imaging errors. On the other hand, with a larger grid size, the amount of information available for reconstruction also grows, resulting in improved reconstruction accuracy. Therefore, to positively impact reconstruction quality with larger grid sizes, the gain in information must outweigh the reduction in imaging quality.

Figure 5b reveals several important trends regarding the performance of different encoding methods. First, HE consistently achieves the highest CNR across all grid sizes compared to alternative methods. Although the CNR for HE declines with increasing grid size, the rate of decline is significantly slower than that observed for RE. This behavior can be attributed to the opposing linear relationship between $\text{SNR}_{\Delta r}$ and $w - \lambda_m$, which remains constant for RE but increases with grid size in the HE approach. Consequently, the slower reduction in imaging error for HE results in a slower decline in reconstruction CNR, underscoring its robustness and suitability for applications requiring high-resolution imaging.

Second, for ROE, a clear improvement in CNR is observed as grid size increases. This outcome arises from dual benefits: enhanced imaging accuracy and greater information gain with larger grids. These factors combine to make ROE a strong contender for high-resolution applications.

Third, RE exhibits a sharp decline in CNR with increasing grid size, even when denoising is applied during the imaging process. This trend aligns with the substantial reduction in imaging error observed for larger grids, as previously discussed, and highlights a clear disadvantage of RE in high-resolution scenarios. Although denoised RE methods offer some improvement over RE, they remain significantly outperformed by both HE and ROE, particularly as grid size increases. This further solidifies the superior performance and robustness of HE and ROE for high-resolution imaging tasks.

\begin{figure}
\centering
\includegraphics[width=1\linewidth]{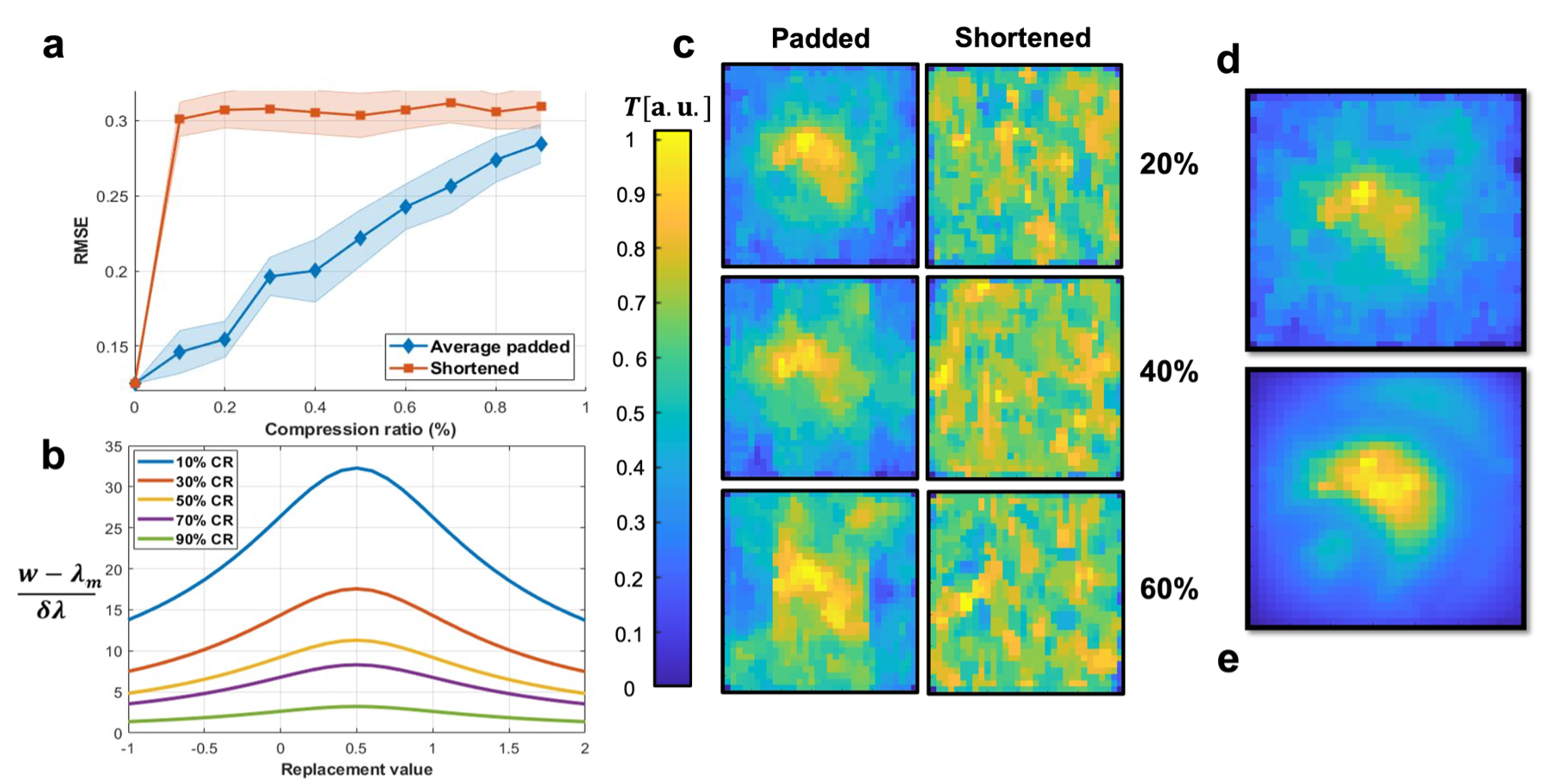}
\caption{\label{fig:Fig6} STEM using compressive sampling. {a}, RMSE vs compression ratio (CR) for both the shortened code and average padding compression techniques. The average padding technique allows to maintain most of the spatial information for lower acquisition times. 
{b}, replacement value vs $SNR_{\delta\lambda}$ for varying compression ratios (CR), a peak can be noticed around $0.5$ for all CRs which aligns with the phenomenological success of the average padding technique. 
{c}, measured intensity image reconstructed using STEM for the different compression methods and varying compression ratios. 
{d}, a STEM image taken with zero compression, i.e., all $n = N$. 
{e}, the ground truth low-light image captured with a CMOS beam profiler.
}
\end{figure}

\subsection{Compressive sampling techniques }

In this section, we would like to understand if images can be captured even for code lengths smaller than the grid size ($n<N$). This feature shares attributes with the concept of compressive sensing for single pixel imaging [29] which allows to capture images with reduced sampling points. Since our goal is to reconstruct $N$ data points using shorter code lengths, we must choose how to generate and analyze them effectively. In order to draw general conclusions regarding the compression abilities of STEM, the modified shorter codes will be generated by randomly removing $m$ points from each sequence in $\mathcal{C}$. 

Resulting in a $(N-m,\tilde{w},\tilde{\lambda}_m,\tilde{\delta\lambda})$ code family $\tilde{\mathcal{C}}$ with different properties from the original family $\mathcal{C}$. A unique modified shorter code will be assigned to each pixel to modulate its intensity and to perform STEM, resulting in a shorter acquisition time. After the image acquisition is performed, the detected signal must be correlated with the different codes to recover an image. 

We will compare two different compressive decoding techniques in this section. The first method involves correlating the detected signal with each pixel’s shortened code from $\tilde{\mathcal{C}}$ to reconstruct its relative intensity. The second technique involves correlating the detected signal which has a length of $N - m$ with the original code family $\mathcal{C}$. However, since the signal was shortened, there is a length mismatch between the codes and the detected signal. 

To mitigate this, the signal is padded with its mean value, exactly at the locations of the removed code entries which will result in an expanded signal of length $N$. To recover the pixels intensity, the expanded signal is correlated with the original code family $\mathcal{C}$ for all pixels. In Figure 6a, the imaging RMSE for HE is presented for varying compression ratios, for the two compressive decompression ratio (CR) is defined to be ratio between the number of removed points ($m$) and the grid size ($N$). It can be noted that the first method fails for all compression ratios while the second method allows to capture most of the spatial information even for high compression rates. A comprehensive explanation of the differences between the two methods and the success of the second technique goes beyond the scope of this article and warrants further investigation. 

However, we highlight a noteworthy observation: the second technique, which involves padding the detected signal with the averaged value, is essentially equivalent to using the original code family $\mathcal{C}$, with each original sequence's entries replaced with a value of $0.5$ precisely at $m$ locations. These $m$ locations correspond to the randomly sampled entries that were removed in the initial step of both compressive sampling techniques.

To understand if the value $0.5$ holds any particular significance, it necessitates a statistical analysis of its effect on $\delta\lambda$. In this article, we provide only numerical evidence for its effectiveness. In Figure 6c, we plot the ratio $(w - \lambda_m)/\delta\lambda$ for different replacement values and for different compression ratios. Notably, there is a prominent peak centered around $0.5$ for all CRs. Since this ratio defines the robustness to code interference noise, the observed peak lends confidence to our choice and partially explains the success of the second compression technique.

\section{Conclusions}

In this paper, we introduced Space-Time Encoded Modulation (STEM), a novel illumination and detection technique for diffuse imaging. STEM allows the attainment of images using single-pixel devices with improved imaging SNR and lower acquisition times. In terms of imaging quality, it was found mathematically that low correlation shifts ($\delta\lambda$) and high weight-to-correlation difference ($w - \lambda_m$) are required for optimal performance, which was further supported in both the synthetic and ex-vivo low-light imaging experiments.

First, we observed that increasing the grid size results in higher imaging errors for the HE and RE cases. This outcome aligns perfectly with our analysis, which indicates that reduced intensity levels on the SLM, denoted as $T_{(i')}$, lead to a linear reduction in the measurement noise SNR ($\text{SNR}_{\delta r}$). For the ROE case, we found that, since it has a non-zero correlation shift $\delta\lambda$, its main noise component is the interference noise. We were able to prove numerically that the interference noise ratio ($\text{SNR}_{\delta\lambda}$) increases for larger code lengths, explaining ROE’s decreasing trend in RMSE for larger grid sizes (Figure 3f).

Second, the Hadamard Encoding (HE) consistently led to the best results across all imaging configurations, including the ex-vivo experiments. HE combines both attributes required for optimal codes: a zero-correlation shift and a high ($w - \lambda_m$) value, which makes it robust to both noise mechanisms. In contrast, the RE approach, while also featuring a zero-correlation shift, demonstrated increased imaging error due to its lower ($w - \lambda_m$) value, rendering it susceptible to measurement noise. This trend persisted in the chicken breast experiments, where RE's performance declined sharply with increasing grid size, particularly under the scattering challenges posed by the heterogeneous medium.

To further demonstrate STEM's capabilities, we extended our analysis to image reconstruction for detecting embedded anomalies using DOI. By employing regularized linearized reconstruction, we demonstrated the reconstruction of anomalies under realistic conditions. Importantly, HE consistently achieved the highest contrast-to-noise ratio across all grid sizes, reaffirming its robustness for diffuse imaging tasks. While CNR for HE declined at larger grid sizes, the reduction was significantly slower compared to RE, owing to HE's superior noise resilience. Additionally, ROE exhibited clear improvement in CNR as grid size increased due to the combined effects of enhanced imaging accuracy and greater information gain, making it a competitive choice for high-resolution applications. In contrast, RE suffered a sharp decline in CNR with increasing grid size, highlighting its limitations for reconstruction tasks.

Finally, we introduced a compressive sampling technique and compared two compression methods. It was observed that utilizing shortened code lengths and padding the detected signal with its averaged values enabled the preservation of spatial information even at high compression ratios. As a partial explanation for this success, our numerical analysis revealed a distinct peak in the $\text{SNR}_{\delta\lambda}$ around 0.5 for varying replacement values and varying compression ratios.

Throughout this paper, several key assumptions and simplifications were made, which can guide future directions and possible extensions.

\begin{enumerate}
    \item A CW system was assumed throughout the paper, due to its simplicity and lower cost. However, future work should focus on implementing STEM for TD operation. TD systems, with their non-linear count rate curves and Poissonian noise characteristics, will require special attention to adapt STEM effectively. We expect these systems to behave as pseudo-CW light sources due to their lower modulation frequencies. Detailed mathematical modeling and careful consideration of these unique features will be essential to fully extend STEM's applicability for state-of-the-art applications.

    \item Throughout the mathematical derivation, we exclusively considered binary codes. This choice allowed us to attain the mathematical bounds for estimating noise components and their effects on the SNR. It's worth mentioning that while SLMs are capable of projecting both binary and non-binary patterns, using a DMD to project binary modulation patterns substantially enhances the system's maximum frame rate. Specifically, a 1-bit encoding can achieve up to 20 kHz, whereas an 8-bit encoding is limited to several hundred Hz. Therefore, by concentrating on binary patterns, we can attain a much higher frame rate and, consequently, a shorter acquisition time.

    \item The STEM method outlined in this paper can also be utilized in other established DOI systems, such as HD-DOT [5], [6], [9]. In HD-DOT, densely packed sources and detectors are arranged over the scattering medium to achieve high-resolution 3D reconstruction of the medium's optical properties. Given that STEM's imaging SNR scales with the number of source and detection points, it is particularly well-suited for high-density imaging scenarios. To demonstrate this, we conducted a detailed simulation study using a realistic breast imaging setup, which is presented in the Supplementary Material section 2.11. The study aimed to validate STEM’s performance under very small source-detector separations and its ability to operate multiple sources simultaneously through temporal modulation. In this study, the light sources were modulated using either Hadamard or Raster encoding schemes and a numerical breast phantom with embedded anomalies was used to investigate the system's behavior under varying signal-to-noise ratios (SNR). The results demonstrated that Hadamard encoding is particularly effective in low-SNR conditions, where it achieved the highest improvement in CNR over Raster encoding. Additionally, Hadamard encoding consistently produced reconstructions with significantly lower variance compared to Raster encoding. Both these attributes strengthen the case for using Hadamard encoding in HD-DOT systems, particularly in scenarios where low-light and high noise is prevalent.

    \item Our measurements were limited to grids of up to 32$\times$32 pixels due to the suboptimal performance of the SLM as an amplitude modulation device. The SLM’s significant losses, caused by the use of linear polarizers and low modulation depth, constrained the signal quality and operable SNR. Its low frame rate and temporal distortions further restricted the operational speed to 2 Hz, making higher-resolution imaging (e.g., 64$\times$64) impractical due to prolonged acquisition times. These limitations can be overcome by using a fast Digital Micromirror Device (DMD), which offers near-ideal modulation depth and frame rates up to 10 kHz, enabling efficient imaging at higher resolutions.
\end{enumerate}

In conclusion, this paper marks the introduction of the innovative concept of Space-Time Encoded Modulation (STEM). We have conducted an in-depth exploration of the optimal operating conditions for STEM, with a particular focus on its applicability to low-light imaging and diffuse imaging scenarios. To provide a comprehensive understanding of STEM, we have developed a detailed mathematical model that not only elucidates its underlying principles but also serves as a valuable guide for its optimization. The insights derived from our mathematical model align with the findings from our experimental work, reinforcing the robustness and efficacy of STEM.

\section{Discussion}

While our study has shed significant light on the potential of STEM in the field of single-pixel imaging, there is still ample room for further investigation and development. Unlocking the full potential of STEM requires ongoing exploration, innovation, and refinement to ensure its applicability and effectiveness across a wider range of imaging scenarios and applications. This promising technology holds the key to advancing the capabilities of single-pixel imaging, and future research endeavors will continue to shape its evolution and impact in the field. 
Moreover, recent studies have demonstrated that deep learning can refine DOI reconstructions or illumination strategies by learning optimal patterns or leveraging post-processing networks [37], [38]. Such approaches could be extended to develop data-driven codes that maximize signal robustness under various noise conditions, complementing our binary pattern framework and potentially improving STEM’s performance in challenging imaging scenarios.
It is important to note that many DOT applications - especially in brain imaging - commonly employ a reflectance geometry, rather than the transmission-based approach considered here. Reflectance-mode brain imaging DOT poses additional challenges due to the multilayered structure of the head which can introduce more complex light propagation paths. Moreover, unlike our transmission-based simulation provided in Supplementary Material section 2.11, brain imaging might involve dynamic changes in optical properties - such as hemodynamic fluctuations - occurring over time. Although the underlying principle of simultaneous source–detector operation remains applicable, these conditions may degrade the performance of STEM if not properly accounted for in the reconstruction algorithms. Consequently, verifying STEM’s effectiveness for reflectance-mode brain imaging requires further investigations and experimentation, including careful calibration of source–detector placements, refined modeling of multilayered tissues, and potential adaptations to handle temporal variations. By addressing these factors, future studies could extend the benefits of STEM - namely high SNR and efficient encoding - to a broader range of DOT applications, including the highly demanding domain of functional brain imaging.

\section*{References}
\begin{enumerate}
    \item D. A. Boas, D. H. Brooks, E. L. Miller, C. A. DiMarzio, M. Kilmer, R. J. Gaudette, and Quan Zhang, “Imaging the body with diffuse optical tomography,” \textit{IEEE Signal Process Mag}, vol. 18, no. 6, pp. 57–75, 2001.
    \item A. H. Hielscher, A. Y. Bluestone, G. S. Abdoulaev, A. D. Klose, J. Lasker, M. Stewart, U. Netz, and J. Beuthan, “Near-Infrared Diffuse Optical Tomography,” \textit{Dis Markers}, vol. 18, no. 5–6, pp. 313–337, 2002.
    \item Y. Hoshi and Y. Yamada, “Overview of diffuse optical tomography and its clinical applications,” \textit{J Biomed Opt}, vol. 21, no. 9, p. 091312, Jul. 2016.
    \item M. L. Altoe, A. Marone, H. K. Kim, K. Kalinsky, D. L. Hershman, A. H. Hielscher, and R. S. Ha, “Diffuse optical tomography of the breast: a potential modifiable biomarker of breast cancer risk with neoadjuvant chemotherapy,” \textit{Biomed Opt Express}, vol. 10, no. 8, p. 4305, Aug. 2019.
    \item M. D. Wheelock, J. P. Culver, and A. T. Eggebrecht, “High-density diffuse optical tomography for imaging human brain function,” \textit{Review of Scientific Instruments}, vol. 90, no. 5, May 2019.
    \item M. Doulgerakis, A. T. Eggebrecht, and H. Dehghani, “High-density functional diffuse optical tomography based on frequency-domain measurements improves image quality and spatial resolution,” \textit{Neurophotonics}, vol. 6, no. 03, p. 1, Aug. 2019.
    \item V. J. Kitsmiller and T. D. O’Sullivan, “Next-generation frequency domain diffuse optical imaging systems using silicon photomultipliers,” \textit{Opt Lett}, vol. 44, no. 3, p. 562, Feb. 2019.
    \item A. Lyons, F. Tonolini, A. Boccolini, A. Repetti, R. Henderson, Y. Wiaux, and D. Faccio, “Computational time-of-flight diffuse optical tomography,” \textit{Nat Photonics}, vol. 13, no. 8, pp. 575–579, Aug. 2019.
    \item M. D. Wheelock, J. P. Culver, and A. T. Eggebrecht, “High-density diffuse optical tomography for imaging human brain function,” \textit{Review of Scientific Instruments}, vol. 90, no. 5, May 2019.
    \item E. M. Frijia et al., “Functional imaging of the developing brain with wearable high-density diffuse optical tomography: A new benchmark for infant neuroimaging outside the scanner environment,” \textit{Neuroimage}, vol. 225, p. 117490, Jan. 2021.
    \item A. K. Fishell et al., “Mapping brain function during naturalistic viewing using high-density diffuse optical tomography,” \textit{Sci Rep}, vol. 9, no. 1, p. 11115, Jul. 2019.
    \item D. Amgar et al., “Resolving the Emission Transition Dipole Moments of Single Doubly Excited Seeded Nanorods via Heralded Defocused Imaging,” \textit{Nano Lett}, vol. 23, no. 12, pp. 5417–5423, Jun. 2023.
    \item D. Du, X. Jin, and R. Deng, “Non-Confocal 3D Reconstruction in Volumetric Scattering Scenario,” \textit{IEEE Trans Comput Imaging}, pp. 1–13, 2023.
    \item A. Farina et al., “Multiple-view diffuse optical tomography system based on time-domain compressive measurements,” \textit{Opt Lett}, vol. 42, no. 14, p. 2822, Jul. 2017.
    \item G. M. Gibson, S. D. Johnson, and M. J. Padgett, “Single-pixel imaging 12 years on: a review,” \textit{Opt Express}, vol. 28, no. 19, p. 28190, Sep. 2020.
    \item A. Basiri et al., “Ultrafast low-pump fluence all-optical modulation based on graphene-metal hybrid metasurfaces,” \textit{Light Sci Appl}, vol. 11, no. 1, p. 102, Apr. 2022.
    \item C. L. Panuski et al., “A full degree-of-freedom spatiotemporal light modulator,” \textit{Nat Photonics}, vol. 16, no. 12, pp. 834–842, Dec. 2022.
    \item F. Rousset et al., “Time-resolved multispectral imaging based on an adaptive single-pixel camera,” \textit{Opt Express}, vol. 26, no. 8, p. 10550, Apr. 2018.
    \item A. Farina et al., “Multiple-view diffuse optical tomography system based on time-domain compressive measurements,” \textit{Opt Lett}, vol. 42, no. 14, p. 2822, Jul. 2017.
    \item S. S. Bawazir et al., “Multiple Access for Visible Light Communications: Research Challenges and Future Trends,” \textit{IEEE Access}, vol. 6, pp. 26167–26174, 2018.
    \item Y. Qiu et al., “Visible Light Communications Based on CDMA Technology,” \textit{IEEE Wirel Commun}, vol. 25, no. 2, pp. 178–185, Apr. 2018.
    \item K. S. Gilhousen et al., “Increased capacity using CDMA for mobile satellite communication,” \textit{IEEE J Sel Areas Commun}, vol. 8, no. 4, pp. 503–514, May 1990.
    \item A. J. Viterbi, \textit{CDMA: Principles of Spread Spectrum Communication}. Addison Wesley Longman, 1995.
    \item H. Marshoud et al., “On the Performance of Visible Light Communication Systems With Non-Orthogonal Multiple Access,” \textit{IEEE Trans Wirel Commun}, vol. 16, no. 10, pp. 6350–6364, Oct. 2017.
    \item S. Arnon et al., Eds., \textit{Advanced Optical Wireless Communication Systems}. Cambridge University Press, 2012.
    \item \textit{Introduction to Fiber-Optic Communications}. Elsevier, 2020.
    \item L. Yin et al., “Performance Evaluation of Non-Orthogonal Multiple Access in Visible Light Communication,” \textit{IEEE Trans Commun}, vol. 64, no. 12, pp. 5162–5175, Dec. 2016.
    \item H. Marshoud et al., “On the Performance of Visible Light Communication Systems With Non-Orthogonal Multiple Access,” \textit{IEEE Trans Wirel Commun}, vol. 16, no. 10, pp. 6350–6364, Oct. 2017.
    \item M. F. Duarte et al., “Single-pixel imaging via compressive sampling,” \textit{IEEE Signal Process Mag}, vol. 25, no. 2, pp. 83–91, Mar. 2008.
    \item D. B. Lindell and G. Wetzstein, “Three-dimensional imaging through scattering media based on confocal diffuse tomography,” \textit{Nat Commun}, vol. 11, no. 1, p. 4517, Sep. 2020.
    \item J. Xia and G. Yao, “Characterizing beef muscles with optical scattering and absorption coefficients in VIS-NIR region,” \textit{Meat Science}, vol. 75, no. 1, p. 78, Jan. 2007.
    \item G. Marquez and S. Thomsen, “Anisotropy in the absorption and scattering spectra of chicken breast tissue,” \textit{Applied Optics}, vol. 37, no. 4, p. 798, Feb. 1998.
    \item R. Yao, X. Intes, and Q. Fang, “Direct approach to compute Jacobians for diffuse optical tomography using perturbation Monte Carlo-based photon ‘replay’,” \textit{Biomed Opt Express}, vol. 9, no. 10, p. 4588, Oct. 2018.
    \item L. I. Rudin, S. Osher, and E. Fatemi, “Nonlinear total variation based noise removal algorithms,” \textit{Physica D}, vol. 60, no. 1–4, pp. 259–268, Nov. 1992.
    \item D. P. Kingma and J. Ba, “Adam: A Method for Stochastic Optimization,” \textit{Int. Conf. Learn. Representations}, Dec. 2014.
    \item J. Yoo et al., “Deep Learning Diffuse Optical Tomography,” \textit{IEEE Trans Med Imaging}, vol. 39, no. 4, pp. 877–887, Apr. 2020.
    \item M. Mozumder et al., “Diffuse optical tomography of the brain: effects of inaccurate baseline optical parameters and refinements using learned post-processing,” \textit{Biomed Opt Express}, vol. 15, no. 8, pp. 4470–4485, Jul. 2024.
    \item N. A. Dang Nguyen et al., “Reconstructing 3D De-Blurred Structures from Limited Angles of View through Turbid Media Using Deep Learning,” \textit{Applied Sciences}, vol. 14, no. 5, Feb. 2024.
\end{enumerate}

\end{document}


\maketitle

\section{Experimental methods}

\subsection{STEM Imaging System and Acquisition}

A continuous-wave (CW) laser source operating at 640~nm with an output power of 50~mW and a beam diameter of 0.8~mm (Becker \& Hickl, BDS-SM-640) was utilized. To expand the beam, an $\times$15 beam expander is employed. The expanded beam is then directed onto an object positioned between two melamine slabs. Subsequently, the scattering pattern was imaged using a lens through a transmissive spatial light modulator (SLM) from Holoeye (LC 2012). The SLM implemented pixel-based temporal modulations at a frame rate of 2~Hz, a selection made to avoid modulation distortions that can arise at higher rates. For controlling the modulations on the SLM, we employed a custom MATLAB program. Following this, the total spatiotemporally modulated light originating from all sub-images was focused onto an amplified, switchable-gain, silicon detector (Thorlabs, PDA36A) operating at a gain of $4.75 \cdot 10^6~\text{V/A}$. The detector's signal was digitized for subsequent analysis using a high-speed A/D converter (Data Translation, DT9832A) with 16-bit resolution and a sampling rate of up to 2~MHz. Data communication with the computer was facilitated through a USB interface.

\subsection{Time Domain DOT System}

A time-domain diffuse optical tomography system was constructed to characterize the melamine foam optical properties. A picosecond laser source (Becker \& Hickl, BDS-SM-640) is used to illuminate the melamine slab with 100~ps pulses at 640~nm and 5~mW average power. On the other side of the sample, the transmitted light is collected using a lens onto an ultra-fast time-resolved hybrid photo-multiplier detector (Becker \& Hickl, HPM-100-06) with 50~ps resolution, 100/s dark count rate and 6~mm cathode diameter. The source and detectors are connected into a time-correlated-single-photon-counting board (Becker \& Hickl, SPC-130-EMN) to measure the photons distribution of time-of-flight (DTOF).

\section{Supplementary Methods}

\subsection{Noise Characterization}

As described in the main text, there are three primary sources of noise in a photodiode. The first one is thermal noise, also known as Johnson noise, which originates from the load resistor. Thermal noise has a spectral density that is frequency-independent, often referred to as white noise. For a receiver with a spectral bandwidth $B$, the mean-square noise current representing the total thermal noise power can be expressed as:

\begin{equation}
\langle i_{\text{th}}^2 \rangle = \sigma_{\text{th}}^2 B = \frac{4k_B T B}{R_L}
\tag{S1}
\end{equation}

Where $R_L$ is the load resistance, $k_B$ is the Boltzmann’s constant, and $T$ is the absolute temperature. In our system $R_L = 50~\Omega$, $T = 25^\circ \text{C}$ and $B = 5~\text{kHz}$. Therefore, we estimate the total thermal noise power in our system to be $\langle i_{\text{th}}^2 \rangle = 1.64 \cdot 10^{-19}~\text{A}^2$.

The next noise source is shot noise, arising from the statistical nature of photodetection. Shot noise can be accurately modeled using a Poisson distribution. However, when the average number of photoelectrons is significantly large ($N_{\text{ph}} \gg 1$), it can be well approximated using Gaussian statistics. The mean-square noise current, or equivalently, the shot noise power, can be represented by the variance of the shot noise current:

\begin{equation}
\langle i_{\text{sn}}^2 \rangle = \sigma_{\text{sn}}^2 B = 2e I_s B = 2e R P_s
\tag{S2}
\end{equation}

Where $e$ is the electron charge, $I_s$ is the signal photocurrent, $R$ is the detector responsivity, and $P_s$ is the signal’s optical power. For our system $I_s \approx 50~\text{nA}$, therefore we estimate the shot noise to be $\langle i_{\text{sn}}^2 \rangle = 8 \cdot 10^{-23}~\text{A}^2$.

The last noise source is dark current noise, which persists even when no light is incident on the photodiode. Similar to shot noise, dark current noise can also be treated as white noise, with noise power given by:

\begin{equation}
\langle i_{\text{dc}}^2 \rangle = \sigma_{\text{dc}}^2 B = 2e I_D B
\tag{S3}
\end{equation}

Where $I_D$ is the dark current of the photodiode. For our system $I_D \approx 1~\text{nA}$, therefore we estimate the dark current noise to be $\langle i_{\text{dc}}^2 \rangle = 1.6 \cdot 10^{-24}~\text{A}^2$. It is evident that the dominant noise source is thermal noise.

It is customary to estimate the noise RMS power to define the Signal-to-Noise Ratio (SNR). The RMS value of thermal current noise is given by $\langle i_{\text{th}}^2 \rangle^{1/2} = 4 \cdot 10^{-10}~\text{A}$.

We additionally offer a visual validation of the normality of the noise by fitting the empirical noise to a Gaussian distribution and contrasting its cumulative distribution function (CDF) with the standard normal CDF using the \texttt{cdfplot()} function in MATLAB. Fig.~S1 demonstrates a significant resemblance between the two distributions and their CDFs, providing further support for our assumption of a normally distributed noise model.

\begin{figure}
\centering
\includegraphics[width=1\linewidth]{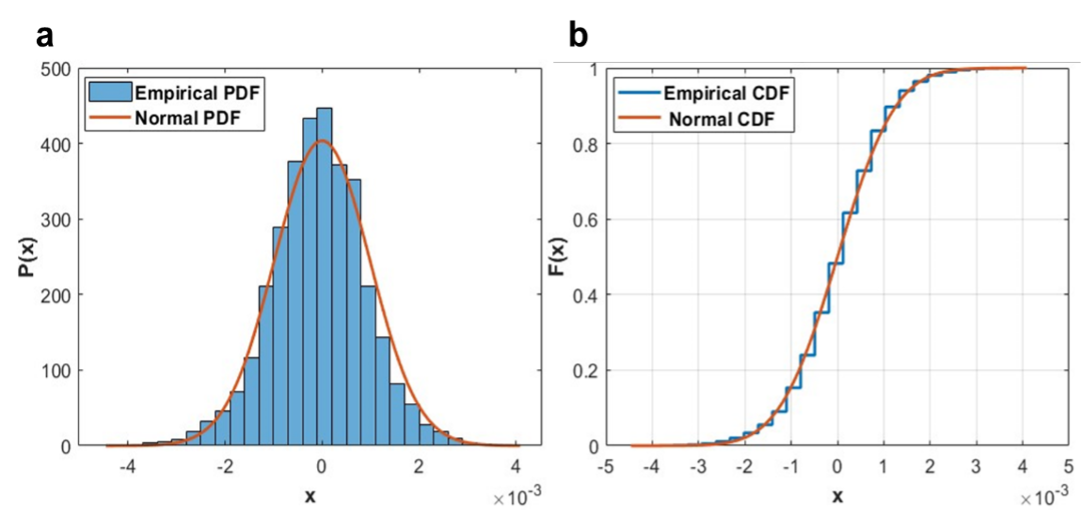}
\captionsetup{labelformat=empty}
\caption{\textbf{Figure S1.} \label{fig:FigS1} Visual normality test. a, Fitting the empirical noise histogram plot to a gaussian distribution. b, Comparison of the empirical cumulative distribution function (CDF) of the normalized measurement noise to the standard normal CDF.}
\end{figure}

\subsection{STEM mathematical derivation}

The experimental setup involves the illumination of a sample containing an embedded object by a continuous wave laser. The objective is to measure the spatially varying transmittance through the sample, denoted as T, using single-pixel detection. Here, T is 2D image comprises of N unknown intensities $T_i$ at pixel i ($N=N_x*N_y$). In the process of forming an image, a lens is utilized to relay the transmittance image into the SLM plane. Subsequently, pixel-dependent binary (0,1) signature codes are applied to the image, where the amplitude of each code is influenced by the intensity of the light on the SLM plane at the specific pixel. Finally, by focusing the light, a detector measures the combined intensity from all coded modulations. The signal at the output of the detector can be written as follows:

\[
r(t)=\sum_{i=1}^N T_i P_i (t)+\delta r(t)\quad(S4)
\]

Where $\delta r(t)$ is the detector noise and $P_i (t)$ is the i’th pixel assigned signature code out of a code family C. An $(n,w,\lambda_m,\delta\lambda)$ signature code is a family of (0,1) sequences of length n, and weight w. The code weight w is the number of ones in the sequence and remains constant across all codes in C. The code size $|C|$ is the number of pixels it can accommodate, i.e., the number of sequences in the family. When the i’th sequence code is displayed on the SLM, it adheres to a specific frame time $T_c$. Where within each frame, a binary on/off key $A_{(i,j)}$ is projected on the SLM screen. Therefore, a general signature code can be written as follows:

\[
P_i (t)=\sum_{j=0}^{n-1} A_{(i,j)} P_{(T_c )} (t-jT_c)\quad(S5)
\]

Where $P_{(T_c )}$ is the rectangular pulse of duration $T_c$ and the total measurement time is denoted as T (i.e., $T=n\cdot T_c$). 

The signature codes’ correlation definitions have undergone several changes compared to the traditional definitions used in multiple access applications. First, the constraint on the signature code’s autocorrelation is removed completely since we only deal with synchronous signals. Second, a new parameter is defined, the mean pairwise cross correlation across all code pairs, $\lambda_m$. Mathematically, it can be written as:

\[
\lambda_m= \frac{1}{|C|^2 - |C|} \sum_{i=1}^{|C|} \sum_{\substack{i' \neq i}} \sum_{j=0}^{n-1} A_{(i,j)} A_{(i',j)}\quad(S6)
\]

Where $A_{(i,j)} A_{(i',j)}$ are the j’th entries from two different codes in C and the coefficient $\frac{1}{|C|^2 - |C|}$ is the number of non-identical code pairs. Unlike the traditional bound on the maximal pairwise cross correlation, we define here a new bound $\delta\lambda$, which is the maximal cross-correlation shift from its mean $\lambda_m$ across all code pairs. Using (3), this bound can be written as:

\[
\sum_{j=0}^{n-1} A_{(i,j)} A_{(i',j)}=\lambda_m+\delta\lambda_{(i,i')} \quad;  \forall i,i' \ (i\neq i')
\quad(S7)
\]
\[
|\delta\lambda_{(i,i')} |\le \delta\lambda \quad;  \forall i,i' \ (i\neq i')
\quad(S8)
\]

Where $\delta\lambda_{(i,i')}$ is $A_i$ and $A_{(i')}$’s specific correlation value shift from $\lambda_m$ which can take any negative or positive value. In essence, rather than limiting the maximum cross-correlation, we express each pairwise cross-correlation as the average value plus a certain shift, and subsequently establish an upper limit for this shift. Using (2), the correlation of two signature sequences integrated over the total measurement time T is given by:

\[
\langle P_i \mid P_{(i')} \rangle =
\int_0^T 
\Bigg(\sum_{j=0}^{n-1} A_{(i,j)} P_{(T_c )} (t-jT_c )\Bigg)
\Bigg(\sum_{k=0}^{n-1} A_{(i',k)} P_{(T_c )} (t-kT_c )\Bigg)
dt
\quad(S9)
\]

By extracting the summation out of the integral and by noting that an integration over two rectangular pulses is always zero when $j\neq k$, we get the following:

\[
\langle P_i \mid P_{(i')} \rangle
=
\sum_{j=0}^{n-1} 
\sum_{k=0}^{n-1}
A_{(i,j)} A_{(i',k)} 
\int_0^T 
P_{(T_c )} (t-jT_c )
P_{(T_c )} (t-kT_c )
dt
=
T_c 
\sum_{j=0}^{n-1} A_{(i,j)} A_{(i',j)}
\quad(S10)
\]

Which can be simplified to the following relations using the code weight definition and cross correlation definition from (4):

\[
\langle P_i \mid P_{(i')} \rangle
=
T_c
\begin{cases}
w, &  i=i' \\
\lambda_m+\delta\lambda_{(i,i')}, & i\neq i'
\end{cases}
\quad(S11)
\]

After defining the temporally integrated correlation of two codes, we can define how the image is formed once the signal is registered. As mentioned, to reconstruct a given pixel intensity value, its corresponding code is to be correlated with the detected signal. For the pixel $i'$, we correlate the received signal $r(t)$ with $P_{(i')} (t)$. Using (1) we attain the following:

\[
I_{(i')}=\langle r \mid P_{(i')} \rangle
=
\sum_{i=1}^N T_i \langle P_i \mid P_{(i')} \rangle
+
\langle \delta r \mid P_{(i')} \rangle
\quad(S12)
\]

By breaking down the summation into two parts: one for $i\neq i'$ and another for $i=i'$, we obtain the following equation:

\[
I_{(i')}=T_c w T_{(i')}+\sum_{i\neq i'} \, T_i \langle P_i \mid P_{(i')} \rangle + \int_0^T \delta r(t) P_{(i')} (t)\, dt
\quad(S13)
\]

It's worth noting that $T_{(i')}$ is the value we aim to reconstruct, hence, naively we could have viewed the first term as our signal, while the second and third terms would be considered as the noise components in our method. However, as we will demonstrate shortly, we can simplify these terms to identify other contributions to our signal. By applying the relationship from (8), we can further decompose the middle term into two components: one corresponding to the constant average cross-correlation value $\lambda_m$ and another related to the pixel-dependent correlation shift $\delta\lambda_{(i,i')}$.

\[
I_{(i')}=T_c w T_{(i')} + \sum_{i\neq i'} T_c \lambda_m T_i + \sum_{i\neq i'} T_c \delta\lambda_{(i,i')} T_i + \int_0^T \delta r(t) P_{(i')} (t)\, dt
\quad(S14)
\]

In the subsequent analysis, we will endeavor to simplify the three last terms to determine their contribution to $T_{(i')}$ and try to establish boundaries for the contributions of the noise sources. First, we define the value M as the total transmittance projected on the SLM (i.e., $M=\sum_i T_i$). This value can be readily determined by measuring the detected signal on the receiver when all the SLM pixels are set to the "ON" position. By using this definition, the first two terms from (11) can be simplified to:

\[
T_c w T_{(i')} + \sum_{i\neq i'} T_c \lambda_m T_i
=
T_c w T_{(i')} + T_c \lambda_m (M-T_{(i')})
=
T_c (w-\lambda_m ) T_{(i')} + M T_c \lambda_m
\quad(S15)
\]

This simplification allowed to acquire a new term proportional to $T_{(i')}$ which will also be affected by the average cross correlation of our code family. Moreover, a constant bias term was found which results in a uniform intensity shift across all $I_{(i')}$ values and will not affect the image. Second, we would like to bound the contribution of the noise stemming from the cross-correlation shift. Using the definition of $\delta\lambda$ from (5) we attain:

\[
\Delta\lambda= \Bigg|\sum_{i\neq i'} T_c \,\delta\lambda_{(i,i')} T_i \Bigg|
\le
T_c \delta\lambda \sum_{i\neq i'} T_i
=
T_c \delta\lambda \left(M - T_{(i')}\right)
\quad(S16)
\]

This upper bound provided us with an estimate for the noise originating from the interference between different code pairs.

\[
\sigma_{\delta\lambda}=T_c \delta\lambda (M-T_{(i')})
\quad(S17)
\]

It's important to emphasize at this point that the non-zero cross-correlation $(\lambda_m\neq0)$ is not the cause of this interference noise. The actual mechanism behind it is the non-uniformity in the pairwise correlation or the non-zero cross correlation shift $(\delta\lambda\neq0)$. Third, we deal with the last term, which is associated with the detector noise. In general, the primary sources of noise in a photodiode can be classified as thermal noise, shot noise, and dark current noise. These noise sources all follow Gaussian statistics when the average number of photoelectrons ($N_{ph}$) and dark photoelectrons ($N_{dark}$) is sufficiently large ($N_{ph},N_{dark}\gg1$). Due to the noise characteristics and following previous works assumptions, we model the noise in the system as an additive white Gaussian noise (AWGN) source with zero mean and variance $\sigma^2$, which incapsulates the combined noise from all sources in the system. We would like to attain an estimate on the variance of the last term from (11). This will allow us to understand its relation to the true signal we aim to measure. Using (2), the last noise term can be written as:

\[
\Delta r=
\sum_{j=0}^{n-1} 
A_{(i',j)}
\int_0^T \delta r(t) P_{(T_c )} (t-jT_c ) \, dt
\quad(S18)
\]

Our noise is modeled as an AWGN, which exhibits stationary statistics, implying that its statistical properties, such as mean and variance, remain constant over time. Consequently, integrating it over non-overlapping time intervals of the same duration yields random variables with identical statistical properties, including the same variance. Additionally, recognizing that $P_{(T_c )}=1$ only for $jT_c<t<(j+1)T_c$ we can modify the integration boundaries as follows:

\[
\Delta r=
\sum_{j=0}^{n-1} 
A_{(i',j)}
\int_0^{T_c} \delta r(t)\, dt
\quad(S19)
\]

This expression represents a summation of the entries $A_{(i',j)}$ multiplied by the integral of the noise term over a duration of $T_c$. Next, we aim to estimate the variance of the integral part from (16). A white Gaussian noise process is a stochastic process where each sample is independently drawn from a Gaussian distribution. When integrated over a time interval $T_c$, the resulting process becomes a random variable, and its variance is given by:

\[
\mathrm{Var}\Big[\int_0^{T_c} \delta r(t)\,dt\Big]
=
\sigma^2 T_c
\quad(S20)
\]

Therefore, we can replace the integral term with a different noise with zero mean and a variance of $\sigma^2 T_c$. Finally, considering that $A_{(i',j)}$ represents a sequence of $n-w$ zeros and w ones, we can conclude that the total noise term $\Delta r$ is essentially a sum of w identical uncorrelated random variables with a zero mean and a variance of $\sigma^2 T_c$. As a result, the total noise variance is given by:

\[
\sigma_r^2=
\mathrm{Var}[\Delta r]
=
\mathrm{Var}\Big[\int_0^T \delta r(t) P_{(i')} (t)\,dt\Big]
=
\sigma^2 T_c w
\quad(S21)
\]

Using (12), (13) and (16), we can identify the total signal as composed of four terms. The true signal, a bias term and the two noise terms.

\[
I_{(i')} = T_c (w-\lambda_m ) T_{(i')} + M T_c \lambda_m + \Delta\lambda + \Delta r
\quad(S22)
\]

Using our prior measurement of M and our knowledge of $T_c$ and $\lambda_m$, the bias term can be removed completely and an estimate for the $i'$-th pixel intensity on the SLM is attained:

\[
\widetilde{(I_{(i')})} 
=
\frac{I_{(i')} - M T_c \lambda_m}{T_c (w-\lambda_m )}
=
T_{(i') }
+
\frac{\Delta\lambda + \Delta r}{T_c (w-\lambda_m )}
\quad(S23)
\]

While we can extract an expression of $T_{(i')}$ it's important to acknowledge the existence of two distinct sources of noise that can individually impact image quality. To assess their impact, we calculate the signal-to-noise ratios (SNR), which quantify the relationship between the desired reconstruction value and the combined influence of both noise sources. By using the bounded interference noise expression from (14) and the estimated measurement noise variance from (18) we can attain the following:

\[
\mathrm{SNR}
=
\frac{T_c (w-\lambda_m ) T_{(i')}}{\sqrt{\sigma_{\delta\lambda}^2 + \sigma_r^2}}
=
\frac{T_c (w-\lambda_m ) T_{(i')}}{\sqrt{\bigl(T_c \delta\lambda(M-T_{(i')})\bigr)^2 + \sigma^2 T_c w}}
\quad(S24)
\]

To understand the individual effect of each noise source and their implication for designing optimal codes, we will distinguish between two cases. The first case is when the interference noise is much larger than the measurement noise $(\sigma_{\delta\lambda}^2 \gg \sigma_r^2)$. This can happen for code families with relatively high correlation shifts. For this case we attain the following SNR:

\[
\mathrm{SNR}_{\delta\lambda}
\approx
\frac{T_c (w-\lambda_m ) T_{(i')}}{T_c \delta\lambda(M-T_{(i')})}
\propto
\frac{w-\lambda_m}{\delta\lambda}
\quad(S25)
\]

We aim to identify SNR dependencies that are not image specific. Hence, we will disregard the $\frac{T_{(i')}}{M-T_{(i')}}$ ratio and exclude it from our analysis. This relation from (22) offers valuable insights for designing and choosing optimal code families. We observe a linear relationship of the SNR with the disparity between w and $\lambda_m$ and an inverse relationship with the maximum correlation shift $\delta\lambda$. Consequently, as explained above, codes that maintain a uniform cross-correlation across all code pairs $(\delta\lambda=0)$ remain resilient to interference noise. The second case is when the measurement noise is much larger than the interference noise $(\sigma_r^2 \gg \sigma_{\delta\lambda}^2)$, which can happen for highly noisy channels. For this case we attain the following SNR:

\[
\mathrm{SNR}_{\delta r}
\approx
\frac{T_c (w-\lambda_m ) T_{(i')}}{\sigma \sqrt{T_c w}}
=
\frac{\sqrt{T_c} (w-\lambda_m ) T_{(i')}}{\sigma \sqrt{w}}
\quad(S26)
\]

Once again, this quantity exhibits proportionality to the difference between w and $\lambda_m$. This reaffirms that, despite the earlier observation, having a low $\lambda_m$ is also necessary to enhance robustness against measurement noise. Moreover, we uncover two other intuitive relationships. First, the SNR displays linear proportionality to the light intensity $T_{(i')}$, aligning with expectations from traditional imaging. Second, the SNR is linearly related to the square root of the frame time $T_c$, mirroring the square root improvement seen in conventional imaging with increasing integration time. Lastly, we observe an inverse relationship between the SNR and the square root of the weight. This underscores the significance of using code familied with a maximized $\frac{(w-\lambda_m)}{\sqrt{w}}$ ratio for optimal performance

\subsection{STEM image reconstruction and preprocessing}

As explained in the main text, STEM can form images by assigning unique coded modulations to different sections of a beam passing through a sample and detecting the light using single-pixel receivers. The image formation algorithm is composed of several steps. Initially, the total projected light on the SLM, denoted as M, is measured independently. To enhance the precision of M estimation, the integration time can be extended as needed until a satisfactory level of accuracy is achieved (Fig. S2 a). Then the STEM image acquisition is performed. First, the raw detected signal is read to the computer (Fig. S2 b). During its acquisition the timestamp of each frame was registered as well and will be used for postprocessing. It was found experimentally, that simply correlating the raw data with the signature code as suggested in equation (9) from the main text leads to poor results. This was due to the temporal shifts the SLM exhibits for above 0.5 Hz frame rates. To avoid these distortions, postprocessing of the raw data was required. The postprocessing step involved looping over the frame time stamps and registering the average detected value for each frame (Fig. S2 c,d). That allowed us to both remove the temporal shifts and to reduce the noise effect due to the averaging step. By using the averaged value of each frame, in order to extract the intensity value $I_{(i')}$, we now need to correlate our detected signal with each code’s entries $A_{(i')} $ and not with the temporal code $P_{(i')} (t)$. Hence, the next step was to correlate the signal with each pixel corresponding signature code to form a vector of length N with the different correlation values of each pixel’s code (Fig. S2 e). To extract the original $T_{(i')}$ values, we use the relation from equation (20) in the main text (Fig. S2 f). To form an image the vector is reshaped such that each entry corresponds to the original pixel location on the SLM itself. 2D median filtering was applied as a postprocessing step (Fig. S2 g).

\begin{figure}
\centering
\includegraphics[width=1\linewidth]{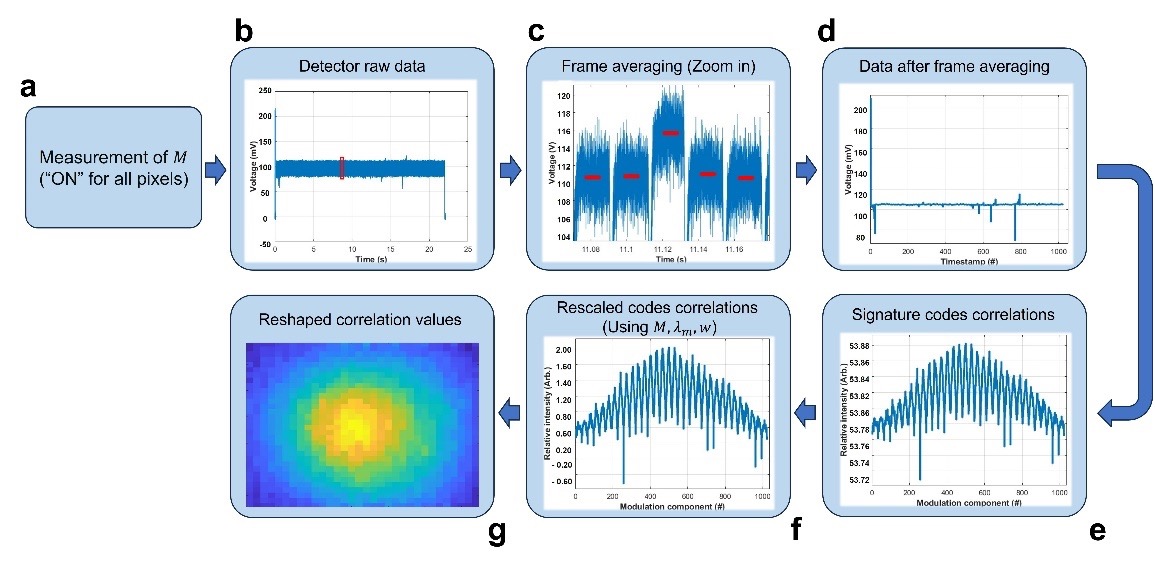}
\captionsetup{labelformat=empty}
\caption{\textbf{Figure S2.} \label{fig:FigS2} STEM image reconstruction. a, measurement of M, performed by using all SLM pixels at “ON” position. b, detector raw data sampled at 10 KHz, we zoom in over the red region. c, zoom in to show the averaging of each frame, the red dots represent the sampled points for averaging. Their x value is their time, and the y value is the average voltage over that frame. d, frames averaged value, it will be correlated with each code’s entries to extract the correlation values. e, relative intensity of each signature code after correlation with the averaged signal. f, rescaling of the correlation results using the values for M, $\lambda_m$ and w g, the reconstrued image after reshaping of the correlation values and applying 2D median filtering.}
\end{figure}

\subsection{Melamine foam characterization}

To reconstruct the embedded objects, an accurate estimation of the background optical properties is required. The absorption $\mu_a$ and reduced scattering coefficients $\mu_s^{\prime}$ of the melamine foam were estimated using an experimental single-pixel time domain DOT system. The sample consisted of two slabs of melamine foam with 2.5 cm width (5 cm total). The optical properties were derived by fitting the measured photon’s distribution of time-of-flight (DTOF) with the expected model given by [1].

\begin{equation}
T(d,t) = \left(4\pi D c\right)^{-1/2} t^{-3/2} e^{-\mu_a c t} \cdot \left\{
\begin{aligned}
&(d - z_0) e^{-(d - z_0)^2 / 4 D c t} - (d + z_0) e^{-(d + z_0)^2 / 4 D c t} \\
+ &(3d - z_0) e^{-(3d - z_0)^2 / 4 D c t} - (3d + z_0) e^{-(3d + z_0)^2 / 4 D c t}
\end{aligned}
\right\}
\label{eq:S27}
\end{equation}

Where $T(d,t)$ is the spatially integrated time resolved transmission through a slab of thickness $d$. $c$ is the speed of light, $D$ is the diffusion coefficient $D = \left\{ 3\left[ \mu_a + \mu_s^{\prime} \right] \right\}^{-1}$ and $z_0$ is the initial scattering depth of the photons $z_0 = \frac{1}{\mu_s^{\prime}}$. By fitting the measured DTOF to the analytical formula, the optical properties $D$ and $\mu_a$ can be extracted. The fitting was done using MATLAB’s curve fitting tool (Fig. S3).

\begin{figure}
\centering
\includegraphics[width=1\linewidth]{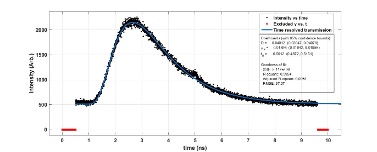}
\captionsetup{labelformat=empty}
\caption{\textbf{Figure S3.} \label{fig:FigS3} Estimating melamine optical properties. The measured DTOF is fitted to the analytical formula [1] to estimate $\mu_a$ and $\mu_s^\prime$.}
\end{figure}

\subsection{Increasing integration time}

As explained in the text, one way to increase the signal SNR is to increase the codes lengths used to differentiate between pixels. Suppose an image $I$ is formed of $N$ unknown intensities $I_i$ with $1 < i < N$. To conduct STEM at least $N$ signature codes must be chosen, one for each $I_i$. In this section we only deal with code lengths $n$ where $n \geq N$. We also assume for simplicity that the length was increased by a ratio $s = n/N$; $s \in {N}$. 

For the HE case, it can be done by taking $N$ rows out of a unipolar Hadamard matrix $H_M$ where $n = sN = 2^M$. By performing this, we generate $N$ distinct codes with weights $w = sN/2$, $\lambda_m = sN/4$ and $\delta \lambda = 0$. The $SNR_{\delta r}$ ratio is given by:

\[
SNR_{\delta r} = \frac{\sqrt{T_c} (w - \lambda_m) T_{i'}}{\sigma \sqrt{w}} = \frac{\sqrt{T_c} (sN/2 - sN/4) T_{i'}}{\sigma \sqrt{sN/2}} = \frac{T_{i'} \sqrt{T_c} \sqrt{N}}{\sigma \sqrt{8}} \sqrt{s}
\tag{S28}
\]

For the RE case, to implement larger code lengths ($n > N$), we simply repeat the original RE code family $s$ times. This is shown below for the case $n = 9$, $N = 3$:

\[
C(n=9, N=3) =
\left(
\begin{array}{ccccccccc}
1 & 0 & 0 & 1 & 0 & 0 & 1 & 0 & 0 \\
0 & 1 & 0 & 0 & 1 & 0 & 0 & 1 & 0 \\
0 & 0 & 1 & 0 & 0 & 1 & 0 & 0 & 1 \\
\end{array}
\right)
\tag{S29}
\]

By performing this, we generate $N$ distinct codes with weights $w = s$, $\lambda_m = 0$ and $\delta \lambda = 0$. The new $SNR_{\delta r}$ improvement ratio for the RE case is given by:

\[
SNR_{\delta r} = \frac{\sqrt{T_c} (w - \lambda_m) T_{i'}}{\sigma \sqrt{w}} = \frac{\sqrt{T_c} (s) T_{i'}}{\sigma \sqrt{s}} = \frac{T_{i'} \sqrt{T_c}}{\sigma} \sqrt{s}
\tag{S30}
\]

This analysis provides an immediate explanation for the results of the low-light imaging experiment (Figure 1b, c). Equations (S28) and (S30) suggest that while both RE and HE exhibit a linear dependency on the square root of $s$, HE offers a distinct SNR slope that is a factor of $\sqrt{N}/\sqrt{8}$ different from RE. Consequently, when $N$ exceeds 8, the HE approach experiences faster improvement compared to its counterpart, a trend that becomes more pronounced as larger grid sizes are employed. Therefore, by leveraging STEM for larger pixel grids, using the Hadamard approach leads to a more efficient way to reduce measurement noise compared to standard methods.

For the ROE case, the code lengths were increased by taking longer random sequences while keeping the weight to be $w = n/2$. The full analysis describing the effect of increasing $n$ on the $(w - \lambda_m)/\sqrt{w}$ ratio for ROE is beyond the scope of this article. However, to fully utilize this approach, further exploration is to be made.

\subsection{Effect of code length for ROE}

From the analysis of low light image acquisition using STEM, it was noticed that in accordance with our mathematical model, HE and RE MSE increases for larger grid sizes. However, for the ROE case, two distinctions were apparent. First, it experiences large imaging errors even for small grid sizes, where the average light powers $T_i$ are relatively high. Second, a decreasing trend in the MSE is noticed for larger grid sizes, contrary to the other methods. We try to explain this phenomenon by noting that since ROE does not exhibit $\delta \lambda = 0$ like its counterparts, it is exposed to the interference noise $\Delta \lambda$ in addition to the regular measurement noise $\Delta r$. Moreover, due to the relatively high MSE for low grid size (where $\Delta r$ is low as well), we can assume that the interference noise is the dominant noise source for ROE.
The decreasing trend in MSE can also be explained if we can prove that the ratio $(w - \lambda_m)/\delta \lambda$ increases for larger code sizes, since it directly governs the $SNR_{\delta \lambda}$ ratio. The full statistical analysis describing the effect of increasing $n$ on $SNR_{\delta \lambda}$ ratio for ROE is beyond the scope of this article. However, by simulating the ROE code generation algorithm for increasing code lengths we can numerically analyze the effect of $n$ on the code correlation properties.
In Fig. S4, we present the outcomes derived from our simulations. Our results reveal that both $(w - \lambda_m)$ and $\delta \lambda$ increase for larger values of $n$, though their rates of increase differ, with $(w - \lambda_m)$ exhibiting a faster growth. Consequently, the ratio $(w - \lambda_m)/\delta \lambda$ also increases as the grid size and code length expand, offering an explanation for the experimentally observed decrease in MSE for larger values of code lengths.

\begin{figure}
\centering
\includegraphics[width=1\linewidth]{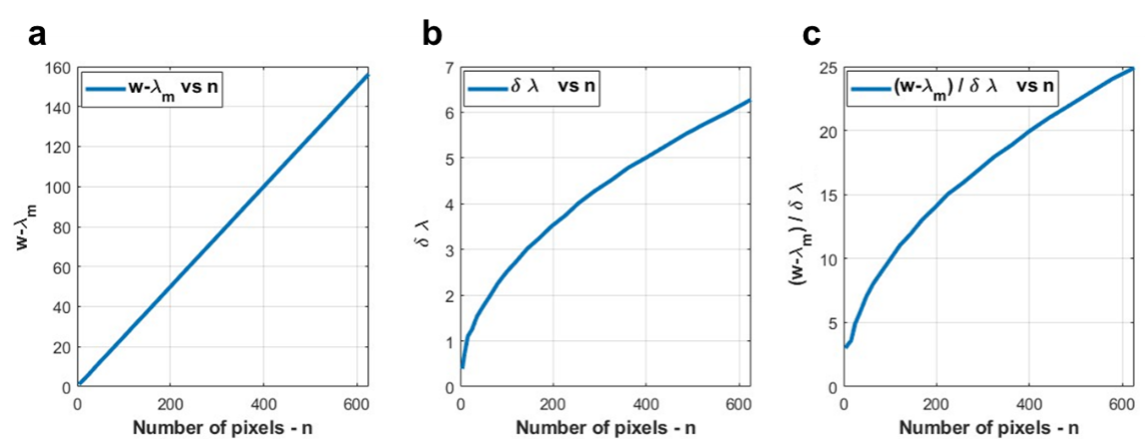}
\captionsetup{labelformat=empty}
\caption{\textbf{Figure S4.} \label{fig:FigS4} ROE correlation vs code length.
{a}, graph of $w - \lambda_m$ vs code length ($n$), as expected increasing $n$ results in a higher number of ones in each code which leads to robustness in terms of interference noise.
{b}, graph of $\delta \lambda$ vs code length, despite the increase in $w - \lambda_m$ for larger $n$’s, we also notice an increasing trend in $\delta \lambda$, which increases the interference noise.
{c}, graph of $\frac{w - \lambda_m}{\delta \lambda}$ vs code length. Although both terms that impact noise from various directions increase, we can discern a rising trend in their ratio for larger values of $n$ (code length). This observation implies that interference robustness increases with the grid size.
}
\end{figure}

\subsection{Denoising methods for raster encoding}

In this section, we provide a comprehensive description and evaluation of the denoising methods applied to the raw RE measurements. These methods were explored to enhance the signal-to-noise ratio (SNR) and validate the performance of the STEM approach. A brief summary of each method is provided below:

\begin{itemize}
    \item \textbf{Gaussian filter} – This method consisted of employing a 2D gaussian filter with $\sigma=0.5$, which was varied to achieve optimal performance.
    
    \item \textbf{Ridge regression} – This method consisted of adding a penalty term $\lambda \|x\|_2^2$ to the least-squares objective and solving for $x$ using: 
    \[
    x = (A^T A + \lambda I)^{-1} A^T b.
    \]
    
    \item \textbf{Tikhonov 1st} – This method consisted of adding a penalty term $\lambda \|L_1 x\|_2^2$ to the least-squares objective where $L_1$ is the first order difference matrix and solving for $x$ using:
    \[
    x = (A^T A + \lambda L_1^T L_1)^{-1} A^T b.
    \]
    
    \item \textbf{Tikhonov 2nd} – This method consisted of adding a penalty term $\lambda \|L_2 x\|_2^2$ to the least-squares objective where $L_2$ is the second order difference matrix and solving for $x$ using:
    \[
    x = (A^T A + \lambda L_2^T L_2)^{-1} A^T b.
    \]
    
    \item \textbf{Iterative Shrinkage-Thresholding Algorithm} – An iterative method for solving $\ell_1$-regularized least-squares problems. At each iteration, it performs a gradient descent step to minimize the least-squares term followed by a soft-thresholding operation to promote sparsity in $x$.
\end{itemize}

The following figure presents the different denoising methods imaging MSE vs. the pixel number and PSNR vs. the integration time.

\begin{figure}
\centering
\includegraphics[width=1\linewidth]{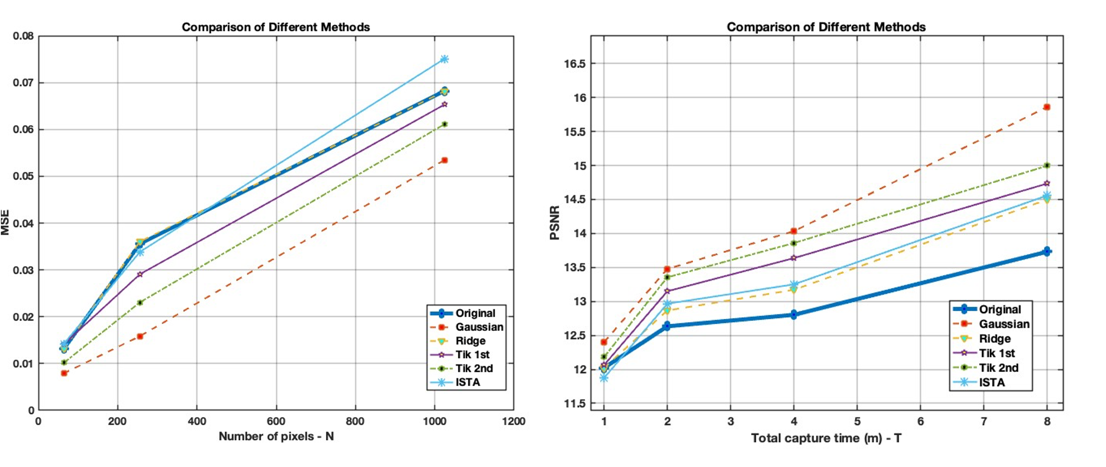}
\captionsetup{labelformat=empty}
\caption{\textbf{Figure S5.} \label{fig:FigS5} a, MSE vs number of pixels for different denoising schemes of the Raster Encoding approach. b, PSNR vs capture time for different denoising schemes for Raster Encoding.
}
\end{figure}

A few observations should be made, regarding the different denoising methods results:

\begin{itemize}
    \item The two top denoising methods are \textbf{Gaussian} and \textbf{Tik 2nd} methods, which outperform all others for all configurations. They are chosen and presented in the main text.
    
    \item \textbf{Ridge regression} has the same result as the non-denoised method for the MSE vs.\ $N$ graph. This is because for regular operation of RE, the sensing matrix $A$ is simply the identity matrix. Hence its solution is 
    \[
    x = (I + \lambda I)^{-1} I b = \frac{b}{1 + \lambda}
    \]
    which when rescaled is exactly $x$ from the non-denoised version. However, when the capture time $T$ increases, $A$ is no longer the identity matrix and the regularization in ridge regression starts to play a significant role. It helps stabilize the solution by reducing the influence of noise and ill-conditioning, resulting in improved performance compared to the non-denoised method.
    
    \item The \textbf{ISTA} method exhibits a non-trivial trend in performance. For smaller grid sizes (e.g., 64 and 128), ISTA initially outperforms the non-denoised method, demonstrating its ability to suppress noise effectively. However, as the grid size increases beyond 256, its performance unexpectedly degrades, falling below that of the non-denoised method. 
     This counterintuitive behavior can be attributed to the high noise levels present at larger grid resolutions. ISTA, being a sparsity-promoting algorithm, applies aggressive regularization, which can oversmooth the solution or suppress critical image details under significant noise. This excessive smoothing reduces the reconstruction quality and leads to poorer results compared to retaining the noisy signal.
    
    Notably, as the integration time increases and noise is progressively suppressed, ISTA regains its advantage, once again achieving better performance than the non-denoised version.
\end{itemize}

\subsection{Retrieval algorithm parameters}

As outlined in the main text, the image reconstruction problem undergoes a linearization process to create a new objective function, as described in Equation~(25). For the purpose of solving this minimization problem, \texttt{PyTorch} was selected as the platform. The image retrieval algorithm starts by initializing the object estimate $\mu^*$ with either zero entries or random values. In each iteration, the estimated measurement vector $m^*$ is computed through a straightforward vector-matrix multiplication involving $H$, along with the calculation of the two norms for the current $\mu^*$. 

Given that we have the true measurement vector $m$ acquired using STEM, \texttt{PyTorch}'s built-in automatic differentiation capabilities are harnessed to efficiently compute gradients for the entire cost function, enabling the optimization of the object estimate $\mu^*$. The optimization process used the parameters from Table~S1 for reconstruction.

\renewcommand{\thetable}{S\arabic{table}} 
\begin{table}[h]
\centering
\caption{parameters for retrieval algorithm}
\vspace{0.5em} 
\begin{tabular}{c c c c c}
\hline
{Number of iterations} & {L2 regularization} & {TV regularization} & {Loss} & {Optimizer} \\
\hline
1e3  & 1e-3           & 1e-1         & MSE             & ADAM \\
\hline
\end{tabular}
\end{table}

\subsection{DOI regularization effect}

As outlined in the main text, the DOT reconstruction phase incorporates two regularization terms: the L2 norm and the TV norm. Figure S6 shows the CNR (defined in the main text) plotted against the number of pixels for all modulation schemes across three scenarios: using both TV and L2 regularization, and each term separately. While an exact solution exists for the case with only L2 regularization, we applied a consistent algorithmic baseline across all scenarios to isolate the effect of each term. Specifically, we used the same reconstruction algorithm, setting the tuning parameter of the term to be ignored to zero.

From Figure S6 we can note several things. TV regularization consistently outperforms L2 in terms of maintaining higher CNR values, demonstrating its effectiveness in preserving contrast and structural details across all modulation schemes. Its trends closely mirror those observed in the combined regularization approach, indicating that TV contributes significantly to the performance when both terms are used. However, despite its lower standalone performance, L2 regularization plays a critical complementary role when combined with TV. The inclusion of L2 in the combined approach leads to the best overall performance, suggesting that its smoothing effect enhances robustness and stability.

\begin{figure}
\centering
\includegraphics[width=1\linewidth]{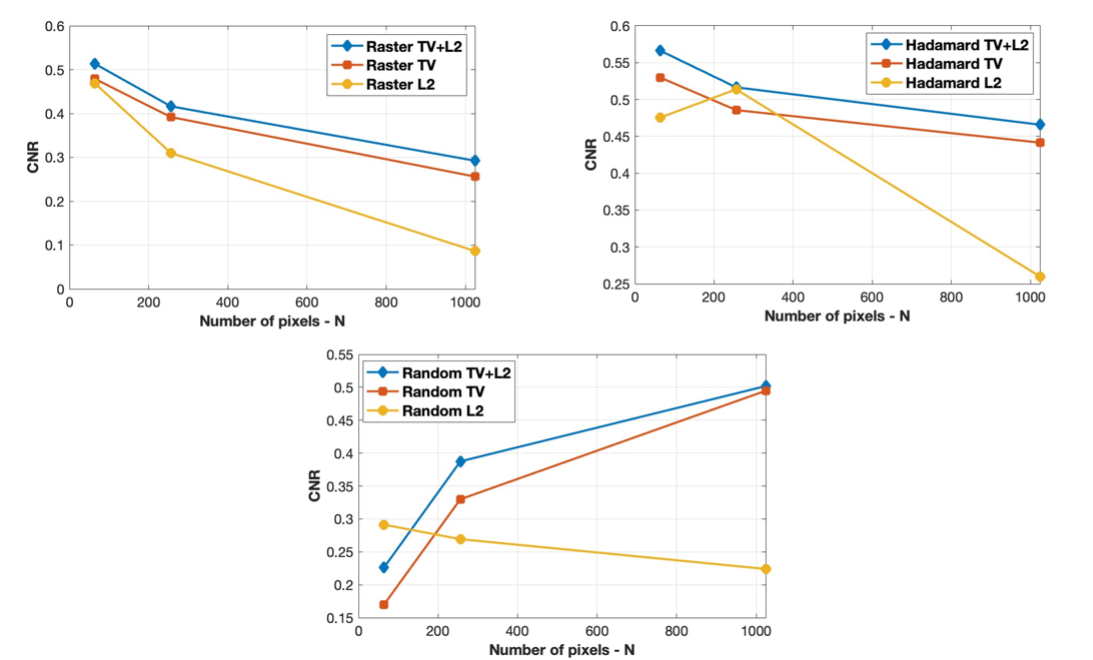}
\captionsetup{labelformat=empty}
\caption{\textbf{Figure S6.} \label{fig:FigS6} Regularization effect on DOI, CNR vs \# of pixels for the three regularization configurations. a, Plot for Raster Encoding. b, Plot for Hadamard Encoding. c, Plot for Random Encoding.}
\end{figure}

\subsection{Computing the Jacobian matrix}

The Jacobian matrix was calculated using perturbative Monte-Carlo with the MCX package [2]. MCX allows to calculate the Jacobian for various imaging scenarios. The object properties can be varied, the detector configuration can be manipulated, and even custom illumination settings can be implemented. We applied the same imaging environment as the experiment. The scattering sample was modeled as a homogenous medium with the same properties as the melamine foam. The source was modeled with the same intensity distribution as in the experiment using the pattern option MCX provides, and the detection area was chosen to collect light from the field of view of the detector in the experiment. The Jacobian matrix is an NxV matrix where N is the number of pixels and V is the number of voxels in the sample. Hence, the i’th row represent the sensitivity of the i’th pixel measurement to a change in any of the voxels in the sample. To calculate the Jacobian, an iterative loop was implemented to loop over all the detector points. In each loop, a constant laser illuminates the sample and light is only registered at a single location, emulating the SLM operation (Fig. S7 a). Hence, we calculate the specific Jacobian for a detector positioned at the location of one pixel on the SLM. After the loop finished, each iteration’s Jacobian was flattened and inserted into the full Jacobian matrix. Resulting in a big matrix representing the sensitivity to a change in the sample for all the different detector/pixel locations (Fig. S7 b).

\begin{figure}
\centering
\includegraphics[width=1\linewidth]{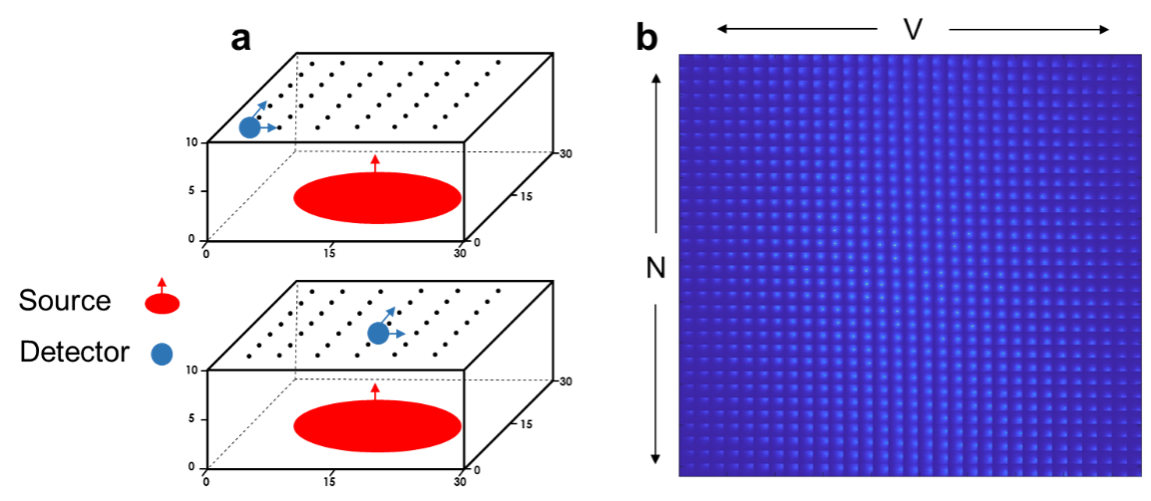}
\captionsetup{labelformat=empty}
\caption{\textbf{Figure S7.} \label{fig:FigS7} Calculating the sensitivity matrix. a, Jacobian calculation loop, in each iteration a constant source was placed and the detector was placed at a different location. The sensitivity was calculated for every iteration’s particular configuration. b, the full sensitivity matrix calculated after iterating over all sources/detectors. It has a size of N x V where N is the grid size and V is the number of voxels in the sample.}
\end{figure}

\subsection{High-Density Diffuse Optical Tomography using STEM – A simulation study}

This section investigates the potential application of the STEM framework for state-of-the-art diffuse optical tomography (DOT) systems. As discussed earlier, STEM enables simultaneous operation of multiple source and detector points, offering a significant signal-to-noise ratio (SNR) improvement compared to traditional sequential scanning approaches. Importantly, the SNR scales with the number of detection and source points used, which makes STEM particularly well-suited for high-density diffuse optical tomography (HD-DOT) applications. In HD-DOT, densely packed sources and detectors are arranged over the scattering medium to achieve high-resolution 3D reconstruction of the medium's optical properties. Hence, we would like to see how well STEM operates in an HD-DOT setup. The data generation pipeline is illustrated in the following figure and was designed to mimic a realistic breast imaging procedure. A numerical breast phantom, developed by Deng et al. (2015) [3] was utilized, accurately representing both the geometry and the heterogeneous optical properties of a real breast. The optical parameters of the phantom correspond to illumination with a laser at a wavelength of 830 nm. To evaluate the system’s ability to detect anomalies, a spherical Gaussian anomaly was embedded within the breast phantom geometry. The anomaly’s optical properties were set to provide a contrast of 2 relative to the mean background optical properties. For this simulation, Over 1000 sources and detectors were evenly distributed along the periphery, with nearest-neighbor separations ranging from 4.5 to 0.5 cm. To evaluate the effect of noise on the imaging performance, additive Gaussian noise was introduced to the detected light signals at varying levels. 
To incorporate the STEM framework into this setup, multiple sources were simultaneously activated, and the combined light was recorded at the detector points. Each source was modulated according to either Hadamard encoding or Raster encoding, as detailed in the main text. In this approach, for each "frame," a specific subset of sources is activated, with the activation pattern dictated by the modulation code. A “1” in the modulation code indicates an active source, while a “0” denotes an inactive source. It is important to emphasize that while Hadamard encoding takes full advantage of simultaneous source operation, Raster encoding corresponds to a sequential scanning approach, where only one source is active at any given frame.

\begin{figure}
\centering
\includegraphics[width=1\linewidth]{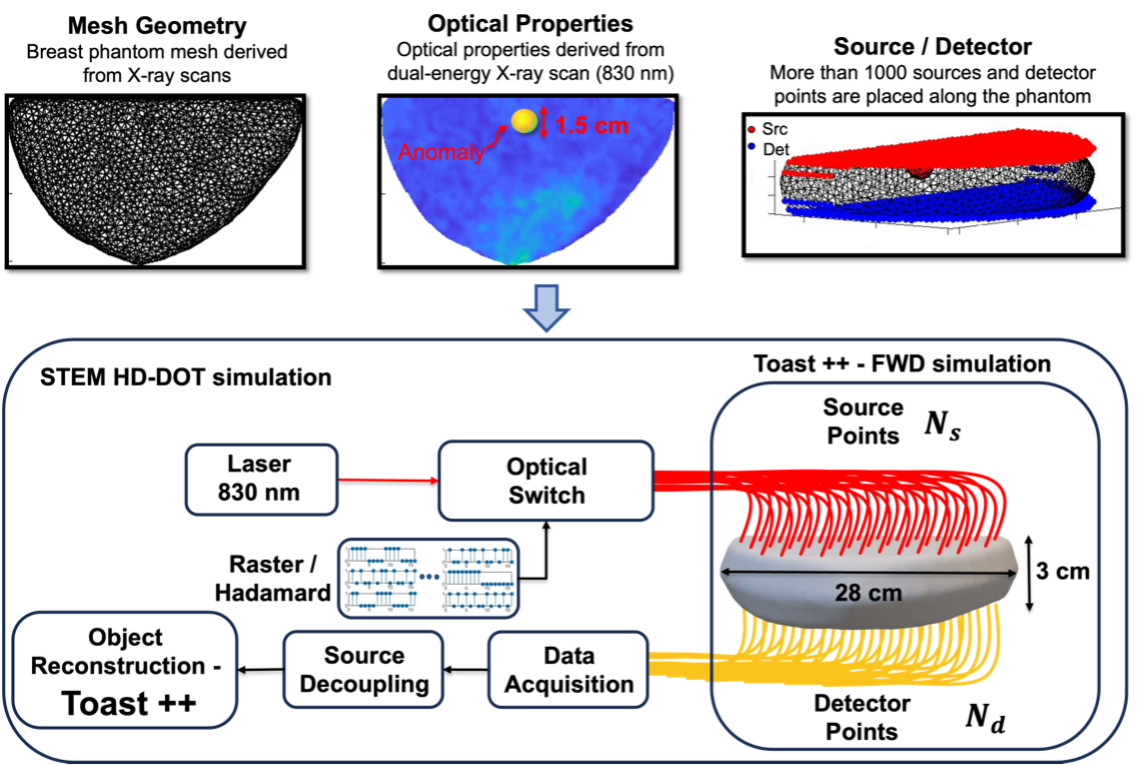}
\captionsetup{labelformat=empty}
\caption{\textbf{Figure S8.} \label{fig:FigS8} STEM HD-DOT configuration. Mesh geometry and optical properties and generated using experimental X-ray scans. Sources and detector are placed along the breast boundary. Finally, Toast++ is used to simulate the forward and backward calculations for employing either Hadamard or Raster encoding techniques to the different source points.}
\end{figure}

Both the forward STEM-HD-DOT simulation and the reconstruction process were conducted using the Toast++ software. The reconstruction procedure consisted of two key steps. First, the measured signals at the detector points were demodulated to extract the contribution of each individual source to the detected signal. Second, the demodulated measurements were processed using the built-in reconstruction algorithms provided by Toast++ to recover the optical properties of the medium. The simulation results are presented in the following figure.

\begin{figure}
\centering
\includegraphics[width=1\linewidth]{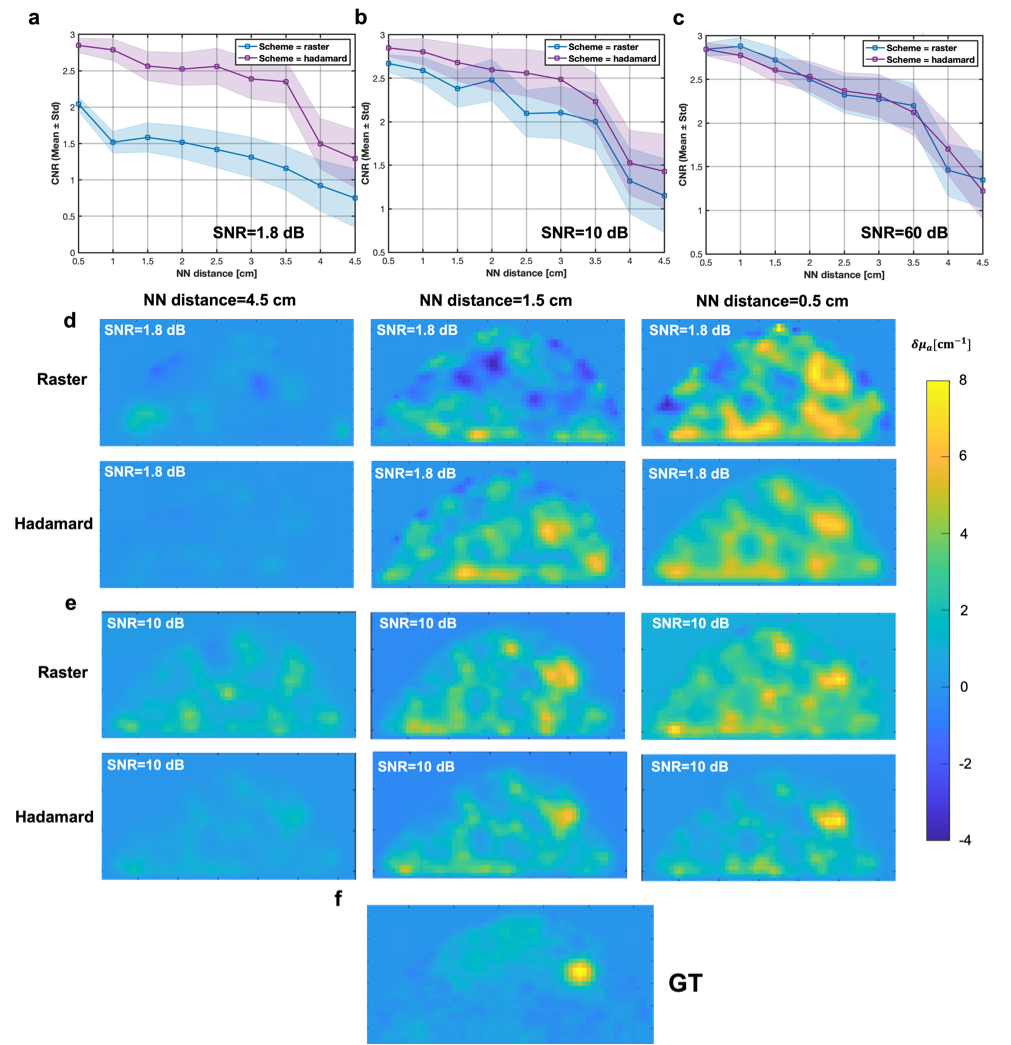}
\captionsetup{labelformat=empty}
\caption{\textbf{Figure S9.} \label{fig:FigS9} STEM HD-DOT results for different signal SNR. a, CNR vs Source density for SNR=1.8 dB for the two schemes. b, CNR vs Density for SNR=10 dB. c, CNR vs Density for SNR=60 dB. d, exemplary reconstruction for the two schemes for different source density levels and for SNR=1.8 dB. e, exemplary reconstruction for the two schemes for different source density levels and for SNR=10 dB.  f, GT optical properties.}
\end{figure}

From Figure S9, we can observe several key trends. First, at low SNR (1.8 dB), STEM with Hadamard encoding demonstrates a substantial improvement in performance compared to traditional Raster encoding. This is particularly crucial for low-light imaging scenarios or environments with significant noise, where signal degradation poses a major challenge. Under these conditions, Hadamard encoding consistently achieves higher contrast-to-noise ratio (CNR) values across all source densities, clearly outperforming Raster encoding. 
As the SNR increases, the performance advantage of Hadamard encoding over Raster encoding diminishes, as illustrated in the figures. At moderate SNR (10 dB), both methods show an overall improvement in CNR, but Hadamard encoding continues to hold a noticeable edge. However, at high SNR (60 dB), the performance difference between Hadamard and Raster encoding becomes negligible. This convergence suggests that the advantages of Hadamard encoding, which arise from its ability to leverage simultaneous activations of multiple sources, are primarily impactful in low-SNR scenarios. 
Further insights can be drawn from the reconstructed images obtained under low-SNR conditions (1.8 dB). As expected, for both Raster and Hadamard encoding, increasing source density enhances anomaly contrast, leading to improved reconstruction resolution and accuracy. However, the primary distinction lies in the reconstructed image noise. Hadamard encoding demonstrates far lower variance in the optical properties compared to Raster encoding across all source densities. This difference underscores Raster encoding's reduced robustness in high-noise systems, resulting in significantly noisier reconstructions.

\section*{References}
\begin{enumerate}
    \item M. S. Patterson, B. Chance, and B. C. Wilson, "Time resolved reflectance and transmittance for the noninvasive measurement of tissue optical properties,"  \textit{Applied Optics}, vol. 28, no. 2331, 1989.
    \item R. Yao, X. Intes, and Q. Fang, "Direct approach to compute Jacobians for diffuse optical tomography using perturbation Monte Carlo-based photon “replay,” \textit{Biomedical Optics Express}, vol. 9, no. 4588, 2018.
    \item Deng, Bin, et al. "Characterization of structural-prior guided optical tomography using realistic breast models derived from dual-energy x-ray mammography." \textit{Biomedical Optics Express}, 2015.
\end{enumerate}